\documentclass{article}
\usepackage{graphicx}
\graphicspath{{Figures/}}
\usepackage{amsmath}
\usepackage{collcell}
\usepackage{colortbl}   % for \cellcolor
\usepackage{geometry}
\usepackage{float}
\usepackage{booktabs,tabularx}
\usepackage[font=footnotesize]{caption}
\usepackage{subcaption}
\usepackage{hyperref}
\usepackage[table]{xcolor}
\usepackage{array}
\usepackage{titlesec}
\usepackage{chemformula}
\usepackage{fourier}
\usepackage[numbers,sort&compress]{natbib}
\usepackage{longtable}

\usepackage{pifont}

\usepackage{collcell,xfp}

\usepackage{adjustbox}
\usepackage{rotating}
\usepackage{authblk}
\usepackage{orcidlink}

\newcommand{\fmtnum}[1]{%
    \ifnum#1<0 %
        \cellcolor{red!20} #1%
    \else
        \ifnum#1<1 %
            \cellcolor{green!20} #1%
        \else
            \cellcolor{blue!20} #1%
        \fi
    \fi
}

\title{Building Trustworthy AI for Materials Discovery: From Autonomous Laboratories to Z-scores}

\author[1]{Benhour Amirian \orcidlink{0009-0004-4053-467X} \footnote{These authors contributed equally to this work.}}
\author[1]{Ashley S. Dale\orcidlink{0000-0001-8233-5258} \textsuperscript{*}\footnote{ashley.dale@utoronto.ca}}
\author[2]{Sergei Kalinin \orcidlink{0000-0001-5354-6152}}
\author[1, 3, 4, 5]{Jason Hattrick-Simpers\orcidlink{0000-0003-2937-3188}\footnote{jason.hattrick.simpers@utoronto.ca}}

\affil[1]{\emph{\small Department of Materials Science and Engineering, University of Toronto, 27 King’s College Cir, Toronto, ON, Canada}}

\affil[2]{\emph{\small Department of Materials Science and Engineering, University of Tennessee, Knoxville, TN 37996, USA}}

\affil[3]{\emph{\small Acceleration Consortium, University of Toronto, 27 King’s College Cir, Toronto, ON, Canada.}}

\affil[4]{\emph{\small Vector Institute for Artificial Intelligence, 661 University Ave, Toronto, ON, Canada.}}

\affil[5]{\emph{\small Schwartz Reisman Institute for Technology and Society, 101 College St, Toronto, ON, Canada.}}

\date{}

\begin{document}

\maketitle

\textbf{ABSTRACT}. Accelerated material discovery increasingly relies on artificial intelligence and machine learning, collectively termed ``AI/ML''. A key challenge in using AI is ensuring that human scientists trust the models are valid and reliable. Accordingly, we define a trustworthy AI framework \textit{GIFTERS} for materials science and discovery to evaluate whether reported machine learning methods are \textit{generalizable}, \textit{interpretable}, \textit{fair}, \textit{transparent}, \textit{explainable}, \textit{robust}, and \textit{stable}. Through a critical literature review, we highlight that these are the trustworthiness principles most valued by the materials discovery community. However, we also find that comprehensive approaches to trustworthiness are rarely reported; this is quantified by a median GIFTERS score of 5/7. We observe that Bayesian studies frequently omit fair data practices, while non-Bayesian studies most frequently omit interpretability. Finally, we identify approaches for improving trustworthiness methods in artificial intelligence and machine learning for materials science by considering work accomplished in other scientific disciplines such as healthcare, climate science, and natural language processing with an emphasis on methods that may transfer to materials discovery experiments. By combining these observations, we highlight the necessity of human-in-the-loop, and integrated approaches to bridge the gap between trustworthiness and uncertainty quantification for future directions of materials science research. This ensures that AI/ML methods not only accelerate discovery, but also meet ethical and scientific norms established by the materials discovery community. This work provides a road map for developing trustworthy artificial intelligence systems that will accurately and confidently enable material discovery. 

\section{Introduction}
Artificial intelligence and machine learning (AI/ML) methods are moving materials discovery towards data-driven approaches that yield more successes with fewer resources~\cite{cheetham2024artificial,hong2020machine,shrestha2024machine,chen2024machine}. AI/ML methods promise to accelerate first principles computations~\cite{zhang2025advancing}, predict experimental process parameters~\cite{guo2024machine}, coordinate high-throughput laboratory experiments~\cite{ahmadi2021machine}, and more. However, sustaining rapid growth in these areas requires human scientists to trust that AI/ML methods are valid, reliable, and correct. 

Trustworthiness in artificial intelligence describes AI/ML models with properties that enable humans to trust the model's predictions and has been identified as an important component of human-machine interaction~\cite{zhou2021understanding, wang2023impact}. Trustworthy artificial intelligence (TAI) frameworks help AI/ML practitioners address trustworthiness in applications such as healthcare~\cite{albahri2023systematic}, criminal justice~\cite{roksandic2022trustworthy}, natural language processing~\cite{ferdaus2024towards}, computer vision~\cite{huang2024pixels}, and economics~\cite{fritz2022financial}. In these cases, TAI frameworks guide the development and implementation of AI by connecting discipline-based methodologies to principles such as \textit{fairness}, \textit{robustness}, and \textit{explainability}. The purpose of this review is to contextualize current best practices within materials science and discovery using a proposed TAI framework for AI/ML methods.

The journey toward TAI began with initiatives such as the European Commission’s High-Level Expert Group on Artificial Intelligence (HLEG AI)~\cite{jobin2019global}, the ``trust comparison index'' proposed by Floridi et al.~\cite{floridi2018ai4people}, and early TAI frameworks~\cite{albinson2019building}. In 2021, the Organization for Economic Co-operation and Development (OECD) formalized principles on artificial intelligence~\cite{canton2021organisation}. These frameworks guide AI systems beyond technical proficiency and towards ethical integrity, transparency, and reliability---qualities that support responsible AI/ML usage in technical domains. 

Proposed frameworks in disciplines beyond materials science vary which aspects of trustworthiness are emphasized. For example, a proposed TAI framework for computer vision identifies transparency, safety and security, privacy, fairness and justice, responsibility, freedom and autonomy, sustainability, and robustness as key elements~\cite{huang2024pixels}. Important elements of TAI for natural language processing include machine ethics, toxicity, privacy, fairness and justice, stereotyping, robustness on adversarial demonstrations, and out-of-distribution robustness~\cite{ferdaus2024towards}. TAI for healthcare includes explainable robotics, predictiveness, explainable decision support, trustworthy review of healthcare, data security, transparency, and trustworthy internet of medical things~\cite{albahri2023systematic}. These technologically-driven frameworks share an emphasis on transparency/explainability/interpretability, social responsibility and fairness, a robustness to attack, and data privacy while emphasizing universal challenges in their respective disciplines. Importantly, these frameworks emphasize principles important to those affected by untrustworthy AI rather than performance metrics. 

Trustworthy AI for materials science and engineering represents stakeholder interests in these algorithms and models. Existing TAI frameworks may be adapted to materials science after acknowledging any unique challenges in the domain and the values of the community surrounding the use of these models, including the alignment of AI/ML models with fundamental scientific laws~\cite{rocken2024accurate, shen2025synergy} and the integration of data with varying levels of fidelity~\cite{chen2021learning, yang2022multi, baird2023materials, fare2022multi}. This review shows that many principles of TAI are already part of common practice in the materials science field in the form of uncertainty quantification (UQ); a longer discussion of the distinction and connection between TAI and UQ is presented in \S \ref{ssec:uq_and_tai}. 

We explicitly clarify that the lack of methods to improve, verify, and explain trustworthiness is not unique to materials science and is shared by other disciplines that leverage AI/ML methods. However, because materials discovery intersects physical sciences, engineering, and data science, the materials discovery community possesses a unique perspective on TAI methods that may contribute towards a unifying theory of trustworthiness within the larger AI/ML community. Accordingly, we summarize the state-of-the-art for TAI practices in materials science by contextualizing existing best practices in the materials discovery community with the first proposed trustworthy AI framework for materials science. 

The review is organized in the following manner: in section \S \ref{ssub:background}, we provide concrete definitions and examples of the proposed framework principles: \textit{generalizability}, \textit{interpretability}, \textit{fairness}, \textit{transparency}, \textit{explainability}, \textit{robustness}, and \textit{stability} (GIFTERS). We next provide case studies for various AI/ML models and applications in materials discovery in \S 
\ref{sec:bayes_models} that highlight the relationship between the TAI framework and UQ in an experimental context. In \S \ref{sec:what_other_people_do}, we summarize efforts towards TAI methods from outside the materials discovery community followed by a discussion in \S \ref{sec:future_directions} of directions where TAI will be expected. We hope that this provides a path towards improved AI/ML methodologies in general, and contributes new directions for AI/ML materials discovery research.

\section{Components of a Trustworthy AI Framework for Materials Science}
\label{ssub:background}

In this section, we propose a trustworthy AI framework \textbf{GIFTERS}, grounded in the values and needs of the materials science community based on trends observed in the literature. The proposed principles of the framework are \textit{Generalizability, Interpretability, Fairness, Transparency, Explainability, Robustness, and Stability}. In practice, these principles frequently overlap in their methodologies. For example, analysis examining a model's interpretability may also provide information about its explainability~\cite{oviedo2022interpretable, minh2022explainable, joyce2023explainable}. A dataset preparation approach that addresses fairness will likely have implications for the model's generalizability~\cite{jadhav2024bias, katare2022bias, maleki2022generalizability}. Accordingly, we emphasize that as the vocabulary surrounding these topics continues to evolve and that depending on usage, there may be overlap, ambiguity, and differences in the naming conventions proposed by various authors cited in this work for their methods.  In this work, principles of GIFTERS are quantified based on apparent author intent for the purpose of emphasizing the methods applied to different principles of GIFTERS trustworthiness. We are grateful for the contributed work in this field, and hope that any re-framing or shift in terminology will not cause offense.

A completely trustworthy AI model may likely---depending on context and use case---require additional principles (e.g., human alignment~\cite{shen2024towards, ji2023ai}, human accountability~\cite{novelli2024accountability}, environmental impact~\cite{bashir2024climate}). There is also an overlap between TAI and the study of ethical AI~\cite{nikolinakos2023ethical}. The narrow scope of the \textit{GIFTERS} framework omits specific ethical concerns and instead connects common methodologies in the materials science community to the proposed framework principles. Our assertion is that any model ignoring the \textit{GIFTERS} principles will fail to achieve trustworthiness and ethicality when \textit{trustworthiness} and \textit{ethicality} are defined in a broader sense than the proposed framework. The above principles are therefore foundational without being comprehensive.

Each component of the framework is described in more detail in the following subsections, and informs case studies in \S \ref{sec:bayes_models} \textit{Evaluating Models Using GIFTERS Framework}. The descriptions are followed by a discussion of how uncertainty quantification supports trustworthy AI by connecting uncertainty quantification methods to trustworthy AI principles. 

\subsubsection*{Generalizability}
\label{sssec:generalization}

A key part of trustworthy AI is generalizability of model performance on unseen data, enabling trust in model predictions when ground-truth values are unknown.  \citet{akrout2023domain} decompose generalization into three properties: first, a model should maintain performance under distribution shifts, it should be able to learn continuously from different scenarios, and it should adapt quickly to larger datasets and increasingly complex computational environments. 

\subsubsection*{Interpretability}
\label{sssec:interpretability}
A model is interpretable when its internal reasoning and behavior (i.e., parameters, architecture, and logical process) can be expressed in human-understandable ways. Interpretability enables trust by providing insight into how predictions are produced through a decision-making process that is meaningful to human practitioners~\cite{oviedo2022interpretable, gao2023interpretability}. Interpretable AI is leverages model design methods such as surrogate modeling~\cite{chatterjee2025acceleration}, selection of different basis kernels~\cite{allen2022machine}, or decision trees~\cite{selvaratnam2023interpretable}. Interpretable methods for non-linear models may leverage feature engineering~\cite{muckley2023interpretable, singh2023augmenting} or symbolic regression~\cite{makke2024interpretable, la2023flexible}. However, most large, non-linear model architectures (including neural networks, convolutional neural networks, and transformers) leverage explainability methods as discussed \textit{vide infra}, sacrificing interpretability for improved performance.

\subsubsection*{Fairness}
\label{sssec:fairness}
A fairly trained model avoids introducing or amplifying data biases, whether those affect individuals~\cite{miron2021evaluating}, groups~\cite{kleinberg2022racial, jain2023awareness}, or, in materials science, classes of materials~\cite{kumagai2022effects, jablonka2020big, merchant2023scaling}. For a comprehensive overview of fairness strategies see~\citet{barocas2023fairness}; proposed techniques for bias mitigation include addressing selection, sampling, and labeling biases, and fair training protocols discussed involve resampling, adversarial learning, and post hoc calibration. 

\subsubsection*{Transparency}
\label{sssec:transparency}
A transparent model provides open access to a model's structure and parameters, training and testing data, and the software for training, testing, and deploying the model. This enables future model users to understand and repeat each step in the development process of the model. Transparency has been recognized as an important issue~\cite{shick2024transparency, wood2024more,van2021open}, and efforts to standardize transparency are presented at length in \S\ref{sec:what_other_people_do}.

\subsubsection*{Explainability}
\label{sssec:explainability}

Explainability enables trust through model predictions that can be understood in the context of input data, typically by implementing a second model to explain the first model. Single instance explainability methods, such as counterfactual explanations~\cite{wellawatte2022model}, are distinct from methods that explain global feature usage such as SHapley Additive exPlanations (SHAP) analysis~\cite{lundberg2017shap}, Local Interpretable Model-agnostic Explanations (LIME) analysis~\cite{ribeiro2016should}, and partial dependence plots~\cite{roy2025prediction}; the latter are ubiquitous within machine learning for materials science~\cite{terashima2023experimental, ravi2023elucidating}.  For a comprehensive overview of different explainability methods, we refer the reader to existing reviews of this topic~\cite{van2021evaluating,oviedo2022interpretable,zhong2022explainable}.

\subsubsection*{Robustness}
\label{sssec:robustness}
Robustness enables trust that the model has learned a meaningful representation of the problem that is not dependent on noise in the data. Within materials science, robustness ensures meaningful predictions in the presence of variations in data due to noise. Noise, the stochastic perturbations that obscure true material properties, is a key aspect of data and well documented within materials science research. Measurement noise arises from experimental imprecision such as shot noise in X-ray diffraction~\cite{slautin2025measurements}, and sampling noise may be introduced by limited or biased dataset curation~\cite{feuer2024select}. Table~\ref{tab:noise_methods} lists the key methods for quantifying different types of noise in materials ML.

\renewcommand{\arraystretch}{1.3}
\setlength{\tabcolsep}{3pt}       

\begin{table}[H]
\centering
\scriptsize
\caption{Methods and tools for quantifying different types of noise in machine learning for materials discovery.}
\label{tab:noise_methods}
\rowcolors{2}{gray!10}{white}
\begin{tabular}{
    >{\centering\arraybackslash}m{2.8cm} 
    >{\centering\arraybackslash}m{4.0cm} 
    >{\centering\arraybackslash}m{5.0cm} 
    >{\centering\arraybackslash}m{2.4cm}
}
\toprule
\rowcolor{gray!25}
\textbf{Type of Noise} & \textbf{Methods/Tools} & \textbf{Description} & \textbf{References} \\
\midrule

Measurement Noise & 
Gaussian process regression (GPyTorch), Bayesian neural networks (PyTorch) & 
Estimate noise in experimental data by modeling it statistically, giving confidence levels for predictions. & 
\cite{gardner2018gpytorch} \\

Sampling noise & 
Stratified sampling, Discrepancy measures (Scikit-learn) & 
Ensure data represents all material types, reducing bias in predictions. & 
\cite{pedregosa2011scikit} \\

Label noise & 
Confident learning (cleanlab), Robust loss functions (PyTorch) & 
Identify and fix wrong labels by checking model agreement or using error-tolerant math. & 
\cite{northcutt2021confident,ghosh2017robust} \\

\bottomrule
\end{tabular}
\end{table}

\subsubsection*{Stability}
\label{sssec:stability}
Given small variations in model hyperparameters or model weights, a stable model will converge to an equivalent final state after training. Stability enables trust that the model results are representative of the data, and not due to selections in hyperparameters. The predictions of a stable model depend more on the input data than on hyperparameters such as random seed, batch size, or learning rate used when training the model. This property provides evidence that the data is featurized and the optimization problem is formulated in a way to promote accurate outcomes~\cite{xiang2021physics}. Within materials science, the requirement for model predictions to be driven by data and not by hyperparameters is motivated by the desire for physically-aligned modeling approaches.

\subsection*{Uncertainty Quantification and Trustworthy AI}
\label{ssec:uq_and_tai}
Uncertainty quantification (UQ) for materials science is a well-developed and continually growing area of research~\cite{shi2025survey, koizumi2024performance}.  To connect UQ to trustworthy AI, it is helpful to briefly review sources of uncertainty and the two existing statistical approaches exist for uncertainty estimation. Bayesian uncertainty incorporates prior physical knowledge, such as crystal symmetries, into posterior distributions \cite{grasselli2025uncertainty}. In contrast, non-Bayesian uncertainty leverages frequentist methods such as ensemble-based variance decomposition that estimate uncertainty through model diversity without relying on explicit priors \cite{allec2025active}, and can introduce flexibility at the cost of interpretability\cite{hirschfeld2020uncertainty}. Finally, UQ estimates are commonly calibrated against an arbitrary statistical model ~\cite{palmer2022calibration, hirschfeld2020uncertainty, psaros2023uncertainty}. 

\begin{figure}[H]
\centering
\captionsetup{font=footnotesize}
\includegraphics[width=0.90\textwidth]{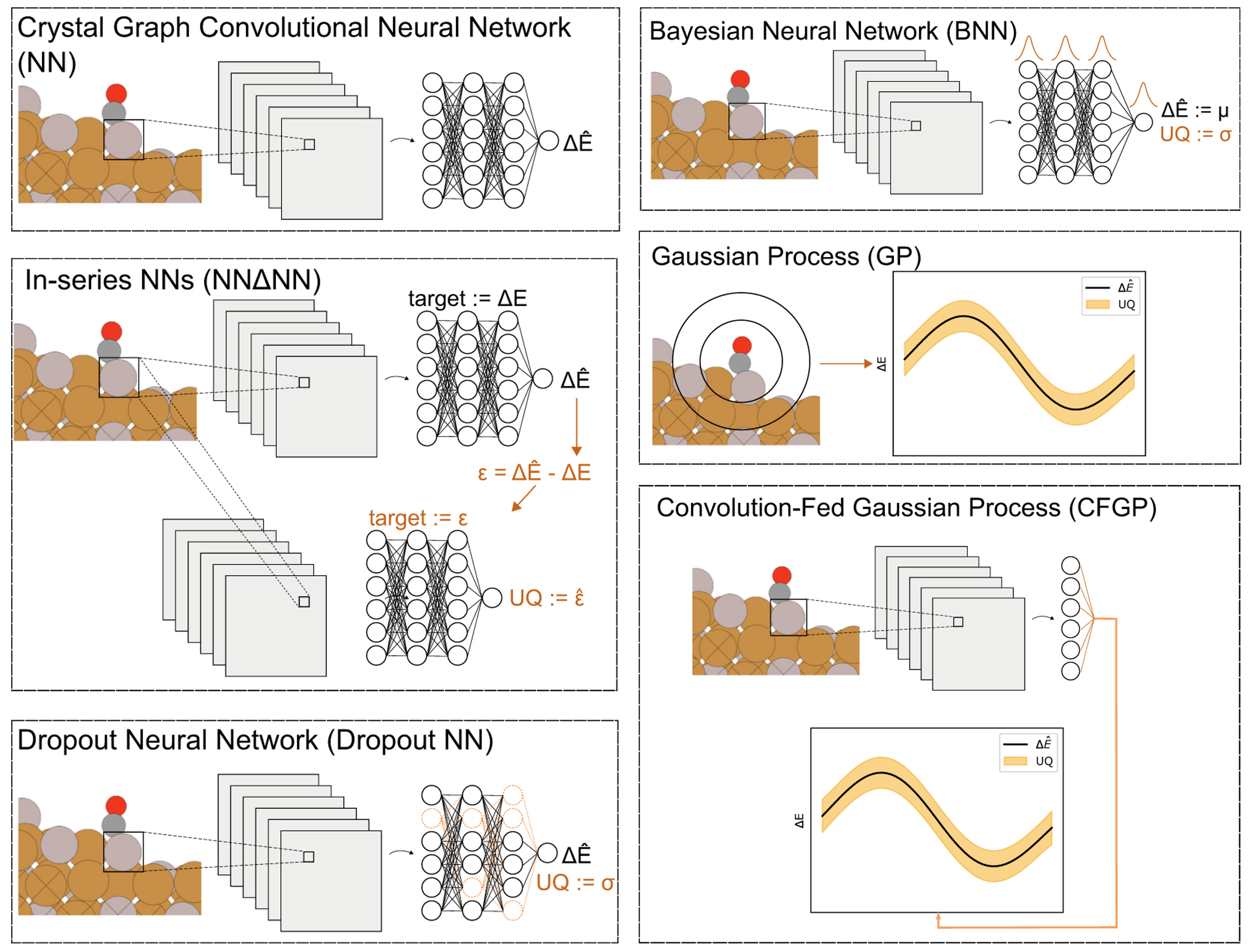}
\caption{Overview of various machine learning-based UQ methods. The figure illustrates six different strategies for predicting DFT-calculated adsorption energies ($\Delta E$) and their associated uncertainties, including a standard crystal graph convolutional neural network (CGCNN), Bayesian neural network, in-series NN (NN$\Delta$NN), Gaussian process, dropout NN, and the CFGP. Each method differs in how it estimates prediction mean ($\mu$) and standard deviation ($\sigma$), with the CFGP achievig the best overall performance in terms of calibration, sharpness, and negative log-likelihood. Reproduced with permission~\cite{tran2020methods}. Copyright 2020 IOP Publishing.}
\label{fig:uncertainity}
\end{figure}

Although UQ is a source of methodologies and metrics, its contribution to trustworthiness originates in how UQ values are calculated, implemented and reported. Different methods of predicting uncertainty are known to result in different uncertainty estimates as visualized in Figure \ref{fig:uncertainity}  \citet{tran2020methods} where a combined modeling approach, convolution-fed Gaussian process (CFGP), outperformed Bayesian neural networks ~\cite{jospin2022hands} and Gaussian process models~\cite{binois2022survey}. When an ensemble model method is used to predict the uncertainty~\cite{rahaman2021uncertainty}, the uncertainty may speak to the stability of the modeling approach rather than uncertainty in the data.  

Finally, there are aspects of TAI that are not connected to established UQ metrology. Specifically, \textit{transparency} specifies access to information about the model and not just access to the model itself. Metrics for software and hyperparameter accessibility exist~\cite{himayat2023software, silva2023quality}, but the social sciences grounding these metrics center human-alignment and stakeholder interests rather than statistical models. Broader principles of TAI may not use any analytical model for evaluating aspects of trust. For example, human accountability may be evaluated in the context of whether model development has followed an established procedure~\cite{xia2024towards}, while environmental impact (e.g., as quantified by energy consumption~\cite{gutierrez2022energy}) may be evaluated using experimental data. As a result, UQ informs and provides a mathematical approach to some aspects of trustworthy AI, but cannot substitute for a trustworthy AI framework. 

\section{Evaluation Using GIFTERS Framework}
\label{sec:bayes_models}

Trustworthy AI/ML for materials discovery is particularly beneficial when experiments are expensive with uncertainty outcomes. Bayesian methods for scientific research minimize uncertainty by combining prior knowledge with new data, then generate predictions with confidence estimates. Non-Bayesian approaches typically emphasize performance, and may consider uncertainty through a separate process such as quantile or heteroscedastic regression for aleatoric noise and deep ensembles or bootstrap resampling for epistemic variance~\cite{sousa2024improving}. In the following section, we evaluate Bayesian and non-Bayesian studies with the GIFTERS framework.

\subsection*{Selection and Scoring Approach}
\begin{figure}[H]
\centering
\captionsetup{font=footnotesize}
\includegraphics[width=0.70\textwidth]{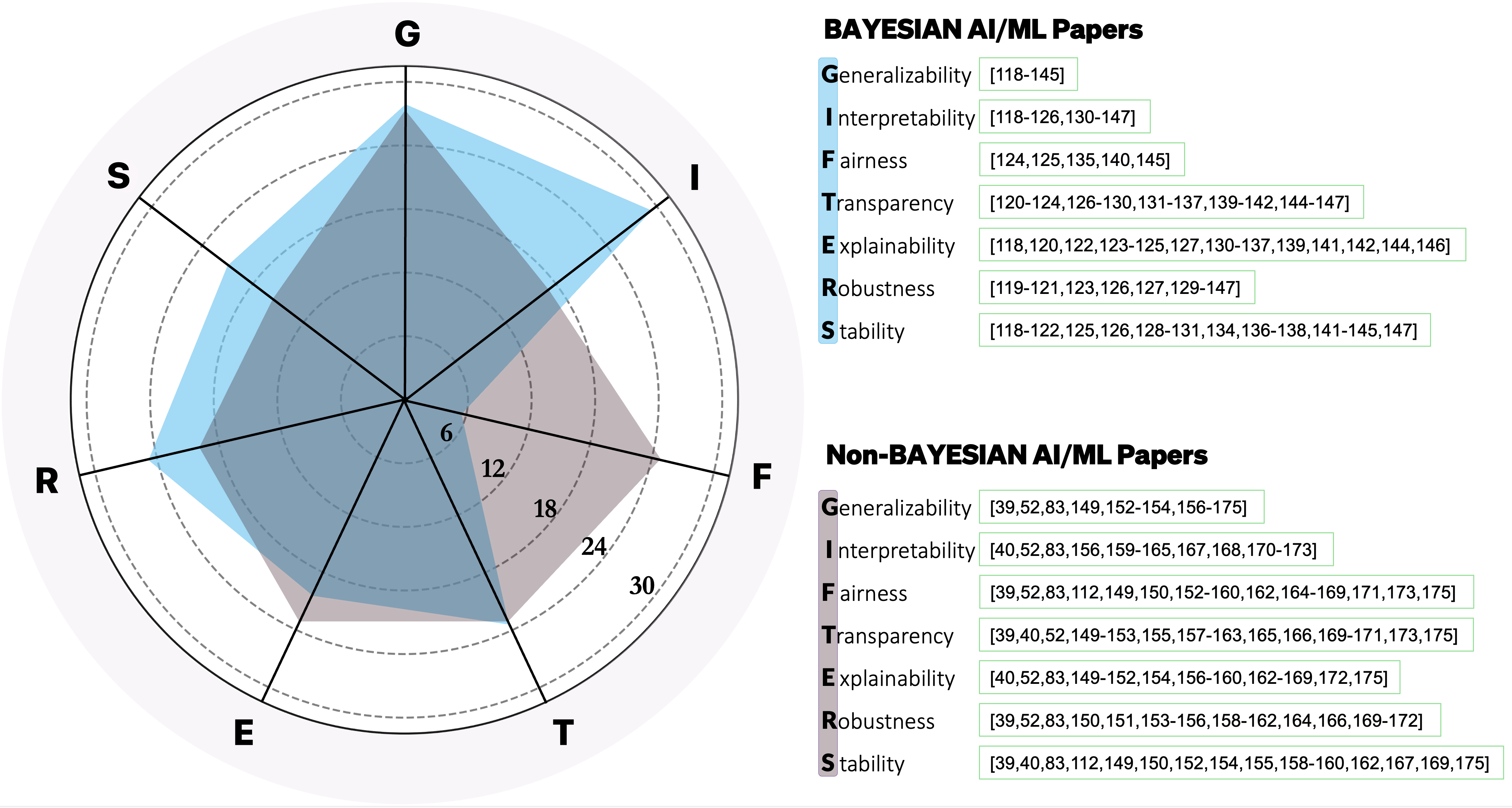}
\caption{GIFTERS-based categorization of trustworthiness attributes for Bayesian and non-Bayesian AI/ML papers: generalizability, interpretability, fairness, transparency, robustness, explainability, and stability.}
\label{fig:tai_by_attribute}
\end{figure}

To examine GIFTERS principles in state-of-the-art research, we examined 63 papers selected for their use of experimental data in model validation, diverse model architectures and algorithms, and the inclusion of both Bayesian~\cite{gongora2020bayesian,battalgazy2025bayesian,khatamsaz2023physics,paulson2019bayesian,gruich2023clarifying,feng2020explainable,adams2024human,kailkhura2019reliable,xu2023small,vandermause2022active,tagade2019attribute,talapatra2018autonomous,startt2024bayesian,makinen2025bayesian,todorovic2019bayesian,zhao2024bayesian,diwale2022bayesian,wakabayashi2022bayesian,shields2021bayesian,wang2023bayesian,liang2021benchmarking,sabanza2025best,de2022experimental,owen2024low,tian2025materials,kusne2020fly,kobayashi2025physics,ray2025refining,chitturi2024targeted,zhong2025towards,hart2024trust} and frequentist approaches to statistics \cite{li2025machine,griffin2020better,koizumi2024performance,muroga2023comprehensive,reinbold2021robust,jones2022impedance,zhang2021leveraging,wang2025experimentally,zheng2025active,tian2024machine,li2023machine,kondo2017microstructure,merchant2023scaling,chatterjee2025acceleration,zhong2022explainable,li2023critical,oviedo2019fast,wang2024data,deng2025machine,wu2023target,drakoulas2024explainable,sodaei2025machine,moon2024active,sarma2022towards,kovacs2023physics,wang2023efficient,zou2024enhanced,hong2022s,mousavi2023modeling,singh2023phase,sharma2025comprehensive,wieczorek2024advancing}. The purpose is to capture a broad picture of how the materials science community uses AI/ML by creating a benchmark analysis of commonly included trustworthiness aspects. Each study was scored (i.e., 0 or 1) against the seven core attributes of trustworthy AI in the GIFTERS framework: generalizability, interpretability, fairness, transparency, explainability, robustness, and stability, as shown in Figure~\ref{fig:tai_by_attribute}. The evaluation results are presented in Table~\ref{table:1}.

\noindent The most commonly discussed attribute in all studies was generalizability where the method for demonstrating generalizability typically relied on selecting a test set with specific properties (e.g. an omitted material group) or preserved performance of a model trained using theoretical data when evaluating experimental data as in ~\cite{merchant2023scaling, li2025machine, wang2025experimentally}. Transparency is typically demonstrated in these works through open, interpretable workflows including model design, training code, and UQ logic; unless a unique method was implemented we omit discussing methods for transparency in the following case studies. Fairness appears least often in Bayesian approaches, with five papers discussing it directly. For non-Bayesian approaches, stability was most often overlooked. In the following section, we present case studies from materials science to explain how each of these attributes is addressed in greater detail.

\begin{table}[h!]
\centering
\caption{GIFTERS scores for every paper reviewed, sorted alphabetically by first author. Items scored ``1'' are shaded for visibility.}
\tiny
\begin{minipage}{0.48\linewidth}
\centering
\begin{tabular}{p{2.8cm}|c|c|c|c|c|c|c|c}
\toprule
\textbf{Reference} & \textbf{G} & \textbf{I} & \textbf{F} & \textbf{T} & \textbf{E} & \textbf{R} & \textbf{S} & \textbf{Score} \\
\hline
\citet{adams2024human} & \cellcolor{green!30}1 & \cellcolor{green!30}1 & 0 & \cellcolor{green!30}1 & \cellcolor{green!30}1 & \cellcolor{green!30}1 & 0 & 5 \\
\citet{battalgazy2025bayesian} & \cellcolor{green!30}1 & \cellcolor{green!30}1 & 0 & 0 & \cellcolor{green!30}1 & 0 & \cellcolor{green!30}1 & 4 \\
\citet{chatterjee2025acceleration} & \cellcolor{green!30}1 & \cellcolor{green!30}1 & \cellcolor{green!30}1 & \cellcolor{green!30}1 & \cellcolor{green!30}1 & \cellcolor{green!30}1 & 0 & 6 \\
\citet{chitturi2024targeted} & \cellcolor{green!30}1 & \cellcolor{green!30}1 & 0 & \cellcolor{green!30}1 & \cellcolor{green!30}1 & \cellcolor{green!30}1 & \cellcolor{green!30}1 & 6 \\
\citet{de2022experimental} & \cellcolor{green!30}1 & \cellcolor{green!30}1 & 0 & \cellcolor{green!30}1 & \cellcolor{green!30}1 & \cellcolor{green!30}1 & \cellcolor{green!30}1 & 6 \\
\citet{deng2025machine}	&\cellcolor{green!30}1	&\cellcolor{green!30}1	&\cellcolor{green!30}1	&\cellcolor{green!30}1	&\cellcolor{green!30}1	&\cellcolor{green!30}1	&\cellcolor{green!30}1	&7 \\
\citet{diwale2022bayesian} & 0 & \cellcolor{green!30}1 & 0 & \cellcolor{green!30}1 & 0 & \cellcolor{green!30}1 & \cellcolor{green!30}1 & 4 \\
\citet{drakoulas2024explainable}	&\cellcolor{green!30}1	&\cellcolor{green!30}1	&\cellcolor{green!30}1	&0	&\cellcolor{green!30}1	&\cellcolor{green!30}1	&0	&5 \\
\citet{feng2020explainable} & 0 &\cellcolor{green!30}1 & 0 &\cellcolor{green!30}1 &\cellcolor{green!30}1 &\cellcolor{green!30}1 & 0 & 4 \\
\citet{gongora2020bayesian} &\cellcolor{green!30}1 &\cellcolor{green!30}1 & 0 & 0 &\cellcolor{green!30}1 & 0 &\cellcolor{green!30}1 & 4 \\
\citet{griffin2020better} & 0 & 0 &\cellcolor{green!30}1 &\cellcolor{green!30}1 &\cellcolor{green!30}1 &\cellcolor{green!30}1 &\cellcolor{green!30}1 & 5 \\
\citet{gruich2023clarifying} &\cellcolor{green!30}1 &\cellcolor{green!30}1 & 0 &\cellcolor{green!30}1 &\cellcolor{green!30}1 & 0 &\cellcolor{green!30}1 & 5 \\
\citet{hart2024trust} & 0 & 0 &\cellcolor{green!30}1 &\cellcolor{green!30}1 &\cellcolor{green!30}1 & 0 & 0 & 3 \\
\citet{hong2022s}	&\cellcolor{green!30}1	&\cellcolor{green!30}1	&\cellcolor{green!30}1	&\cellcolor{green!30}1	&0	&\cellcolor{green!30}1	&0	&5 \\
\citet{jones2022impedance} &\cellcolor{green!30}1 & 0 &\cellcolor{green!30}1 &\cellcolor{green!30}1 &\cellcolor{green!30}1 & 0 &\cellcolor{green!30}1 & 5 \\
\citet{kailkhura2019reliable} &\cellcolor{green!30}1 &\cellcolor{green!30}1 &\cellcolor{green!30}1 &\cellcolor{green!30}1 &\cellcolor{green!30}1 & 0 & 0 & 5 \\
\citet{khatamsaz2023physics} &\cellcolor{green!30}1 &\cellcolor{green!30}1 & 0 &\cellcolor{green!30}1 &\cellcolor{green!30}1 &\cellcolor{green!30}1 &\cellcolor{green!30}1 & 6 \\
\citet{kondo2017microstructure} &\cellcolor{green!30}1 & 0 &\cellcolor{green!30}1 &\cellcolor{green!30}1 &\cellcolor{green!30}1 &\cellcolor{green!30}1 &\cellcolor{green!30}1 & 6 \\
\citet{kobayashi2025physics} &\cellcolor{green!30}1 &\cellcolor{green!30}1 & 0 &\cellcolor{green!30}1 &\cellcolor{green!30}1 &\cellcolor{green!30}1 &\cellcolor{green!30}1 & 6 \\
\citet{koizumi2024performance} & 0 & 0 &\cellcolor{green!30}1 & 0 & 0 & 0 &\cellcolor{green!30}1 & 2 \\
\citet{kovacs2023physics}	&\cellcolor{green!30}1	&\cellcolor{green!30}1	&\cellcolor{green!30}1	&0	&\cellcolor{green!30}1	&0	&0	&4 \\
\citet{kusne2020fly} &\cellcolor{green!30}1 &\cellcolor{green!30}1 & 0 &\cellcolor{green!30}1 &\cellcolor{green!30}1 &\cellcolor{green!30}1 &\cellcolor{green!30}1 & 6 \\
\citet{li2023critical}	&\cellcolor{green!30}1	&\cellcolor{green!30}1 &\cellcolor{green!30}1 &\cellcolor{green!30}1 &\cellcolor{green!30}1 &\cellcolor{green!30}1 &\cellcolor{green!30}1 & 7 \\
\citet{li2023machine} &\cellcolor{green!30}1 & 0 &\cellcolor{green!30}1 &\cellcolor{green!30}1 &\cellcolor{green!30}1 & 0 & 0 & 4 \\
\citet{li2025machine}   &\cellcolor{green!30}1 & 0 &\cellcolor{green!30}1 &\cellcolor{green!30}1 &\cellcolor{green!30}1 & 0 &\cellcolor{green!30}1 & 5 \\
\citet{liang2021benchmarking} &\cellcolor{green!30}1 &\cellcolor{green!30}1 & 0 &\cellcolor{green!30}1 &\cellcolor{green!30}1 &\cellcolor{green!30}1 &\cellcolor{green!30}1 & 6 \\
\citet{makinen2025bayesian} &\cellcolor{green!30}1 &\cellcolor{green!30}1 & 0 & 0 &\cellcolor{green!30}1 &\cellcolor{green!30}1 &\cellcolor{green!30}1 & 5 \\
\citet{merchant2023scaling} &\cellcolor{green!30}1 & 0 &\cellcolor{green!30}1 &\cellcolor{green!30}1 & 0 &\cellcolor{green!30}1 &\cellcolor{green!30}1 & 5 \\
\citet{moon2024active}	&\cellcolor{green!30}1	&0	&\cellcolor{green!30}1	&\cellcolor{green!30}1	&\cellcolor{green!30}1	&\cellcolor{green!30}1	&0	&5 \\
\citet{mousavi2023modeling}	&\cellcolor{green!30}1	&\cellcolor{green!30}1	&0	&0	&\cellcolor{green!30}1	&\cellcolor{green!30}1	&0	&4 \\
\citet{muroga2023comprehensive} & 0 & 0 & 0 &\cellcolor{green!30}1 &\cellcolor{green!30}1 &\cellcolor{green!30}1 & 0 & 3 \\
\multicolumn{9}{c}{\textbf{...}} \\ % <-- continuation marker
\end{tabular}
\end{minipage}%
\hspace{3pt}
\begin{minipage}{0.48\linewidth}
\centering
\begin{tabular}{p{2.8cm}|c|c|c|c|c|c|c|c}
\multicolumn{9}{c}{\textbf{...}} \\ % <-- continuation marker
\citet{owen2024low} &\cellcolor{green!30}1 &\cellcolor{green!30}1 & 0 &\cellcolor{green!30}1 &\cellcolor{green!30}1 &\cellcolor{green!30}1 &\cellcolor{green!30}1 & 6 \\
\citet{oviedo2019fast}	&\cellcolor{green!30}1	&\cellcolor{green!30}1	&\cellcolor{green!30}1	&\cellcolor{green!30}1	&\cellcolor{green!30}1	&\cellcolor{green!30}1	&\cellcolor{green!30}1	&7 \\
\citet{ray2025refining} &\cellcolor{green!30}1 &\cellcolor{green!30}1 & 0 &\cellcolor{green!30}1 &\cellcolor{green!30}1 &\cellcolor{green!30}1 &\cellcolor{green!30}1 & 6 \\
\citet{reinbold2021robust} & 0 &\cellcolor{green!30}1 & 0 &\cellcolor{green!30}1 &\cellcolor{green!30}1 & 0 &\cellcolor{green!30}1 & 4 \\
\citet{sabanza2025best} &\cellcolor{green!30}1 &\cellcolor{green!30}1 & 0 &\cellcolor{green!30}1 &\cellcolor{green!30}1 &\cellcolor{green!30}1 &\cellcolor{green!30}1 & 6 \\
\citet{sarma2022towards}	&\cellcolor{green!30}1	&\cellcolor{green!30}1	&\cellcolor{green!30}1	&0	&\cellcolor{green!30}1	&0	&\cellcolor{green!30}1	&5 \\
\citet{sharma2025comprehensive}	&\cellcolor{green!30}1	&0	&0	&0	&0	&0	&0	&1 \\
\citet{shields2021bayesian} &\cellcolor{green!30}1 &\cellcolor{green!30}1 & 0 &\cellcolor{green!30}1 &\cellcolor{green!30}1 &\cellcolor{green!30}1 &\cellcolor{green!30}1 & 6 \\
\citet{singh2023phase}	&\cellcolor{green!30}1	&\cellcolor{green!30}1	&\cellcolor{green!30}1	&\cellcolor{green!30}1	&0	&0	&0	&4 \\
\citet{sodaei2025machine}	&\cellcolor{green!30}1	&\cellcolor{green!30}1	&\cellcolor{green!30}1	&\cellcolor{green!30}1	&\cellcolor{green!30}1	&0	&0	&5 \\
\citet{startt2024bayesian} &\cellcolor{green!30}1 &\cellcolor{green!30}1 & 0 & 0 & 0 &\cellcolor{green!30}1 &\cellcolor{green!30}1 & 4 \\
\citet{talapatra2018autonomous} &\cellcolor{green!30}1 &\cellcolor{green!30}1 & 0 &\cellcolor{green!30}1 & 0 & 0 &\cellcolor{green!30}1 & 4 \\
\citet{tagade2019attribute} &\cellcolor{green!30}1 & 0 & 0 &\cellcolor{green!30}1 &\cellcolor{green!30}1 &\cellcolor{green!30}1 & 0 & 4 \\
\citet{tian2024machine} &\cellcolor{green!30}1 &\cellcolor{green!30}1 &\cellcolor{green!30}1 & 0 &\cellcolor{green!30}1 &\cellcolor{green!30}1 & 0 & 5 \\
\citet{tian2025materials} &\cellcolor{green!30}1 &\cellcolor{green!30}1 & 0 &\cellcolor{green!30}1 &\cellcolor{green!30}1 &\cellcolor{green!30}1 &\cellcolor{green!30}1 & 6 \\
\citet{todorovic2019bayesian} &\cellcolor{green!30}1 &\cellcolor{green!30}1 & 0 &\cellcolor{green!30}1 &\cellcolor{green!30}1 &\cellcolor{green!30}1 &\cellcolor{green!30}1 & 6 \\
\citet{vandermause2022active} &\cellcolor{green!30}1 &\cellcolor{green!30}1 & 0 &\cellcolor{green!30}1 & 0 &\cellcolor{green!30}1 &\cellcolor{green!30}1 & 5 \\
\citet{wakabayashi2022bayesian} &\cellcolor{green!30}1 &\cellcolor{green!30}1 & 0 &\cellcolor{green!30}1 &\cellcolor{green!30}1 &\cellcolor{green!30}1 & 0 & 5 \\
\citet{wang2023bayesian} &\cellcolor{green!30}1 &\cellcolor{green!30}1 &\cellcolor{green!30}1 &\cellcolor{green!30}1 &\cellcolor{green!30}1 &\cellcolor{green!30}1 &\cellcolor{green!30}1 & 7 \\
\citet{wang2023efficient}	&\cellcolor{green!30}1	&0	&\cellcolor{green!30}1	&\cellcolor{green!30}1	&\cellcolor{green!30}1	&\cellcolor{green!30}1	&\cellcolor{green!30}1	&6 \\
\citet{wang2024data}	&\cellcolor{green!30}1	&\cellcolor{green!30}1	&0	&\cellcolor{green!30}1	&0	&\cellcolor{green!30}1	&0	&4 \\
\citet{wang2025experimentally} &\cellcolor{green!30}1 & 0 &\cellcolor{green!30}1 & 0 &\cellcolor{green!30}1 &\cellcolor{green!30}1 &\cellcolor{green!30}1 & 5 \\
\citet{wieczorek2024advancing}	&\cellcolor{green!30}1	&0	&\cellcolor{green!30}1	&\cellcolor{green!30}1	&\cellcolor{green!30}1	&0	&\cellcolor{green!30}1	&5 \\
\citet{wu2023target}	&\cellcolor{green!30}1	&\cellcolor{green!30}1	& 0	&\cellcolor{green!30}1	&\cellcolor{green!30}1	& 0 & 0	&4 \\
\citet{xu2023small} &\cellcolor{green!30}1 &\cellcolor{green!30}1 &\cellcolor{green!30}1 & 0 &\cellcolor{green!30}1 & 0 &\cellcolor{green!30}1 & 5 \\
\citet{zhang2021leveraging} &\cellcolor{green!30}1 & 0 &\cellcolor{green!30}1 &\cellcolor{green!30}1 & 0 &\cellcolor{green!30}1 & 0 & 4 \\
\citet{zhao2024bayesian} &\cellcolor{green!30}1 &\cellcolor{green!30}1 & 0 &\cellcolor{green!30}1 &\cellcolor{green!30}1 &\cellcolor{green!30}1 &\cellcolor{green!30}1 & 6 \\
\citet{zheng2025active} & 0 & 0 &\cellcolor{green!30}1 &\cellcolor{green!30}1 & 0 &\cellcolor{green!30}1 &\cellcolor{green!30}1 & 4 \\
\citet{zhong2022explainable}	&\cellcolor{green!30}1	&\cellcolor{green!30}1	&\cellcolor{green!30}1	&0	&\cellcolor{green!30}1	&\cellcolor{green!30}1	&\cellcolor{green!30}1	&6 \\
\citet{zhong2025towards} &\cellcolor{green!30}1 &\cellcolor{green!30}1 & 0 &\cellcolor{green!30}1 &\cellcolor{green!30}1 &\cellcolor{green!30}1 &\cellcolor{green!30}1 & 6 \\
\citet{zou2024enhanced}	&\cellcolor{green!30}1	&\cellcolor{green!30}1	&0	&\cellcolor{green!30}1	&0	&\cellcolor{green!30}1	&0	&4 \\
\hline
Average & 0.8709 & 0.7258 & 0.4677 & 0.7742 & 0.7903 & 0.6935 & 0.6451 & 4.967 \\
\bottomrule
\end{tabular}
\end{minipage}
\label{table:1}
\end{table}

\subsection*{Materials Discovery Case Studies}

\begin{figure}[H]
\centering
\captionsetup{font=footnotesize}
\includegraphics[width=0.85\textwidth]{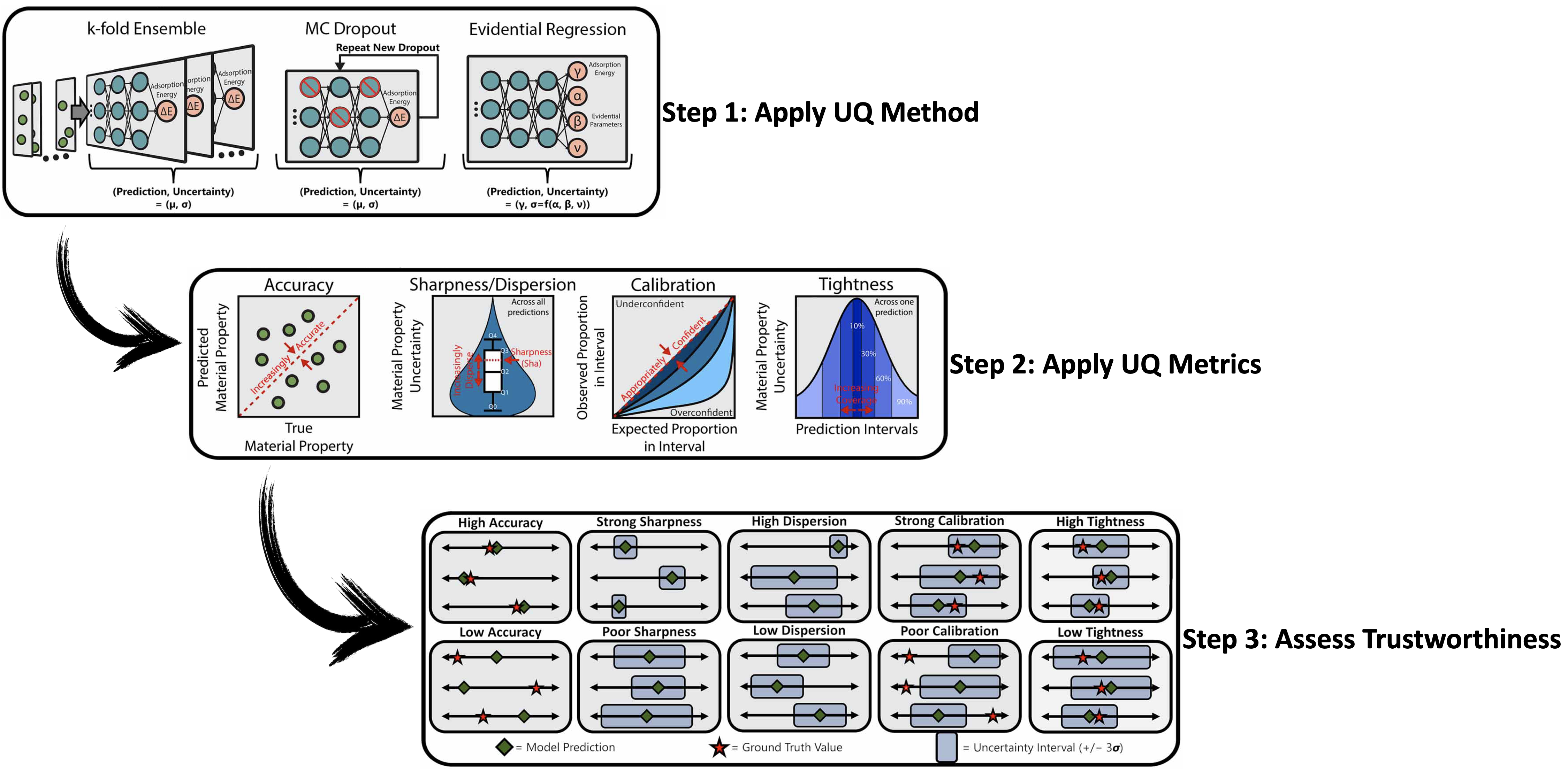}
\caption{Three-step uncertainty-aware workflow. Step 1 applies one of three UQ modules, k-fold ensemble, Monte Carlo dropout, or evidential regression, to generate predictions with error bars. Step 2 scores those outputs with accuracy, sharpness, dispersion, calibration, and tightness metrics. Step 3 charts high versus low trust cases, giving users an assessment of model reliability. Reproduced with permission~\cite{gruich2023clarifying}. Copyright 2023 IOP Publishing.}
\label{fig:uq_pipeline}
\end{figure}

We begin with a study that emphasizes trustworthiness through a structured approach to uncertainty-aware prediction in catalysis using Bayesian neural networks \cite{gruich2023clarifying}. This work benchmarks three UQ methods---k-fold ensembling \cite{baumann2014reliable}, Monte Carlo dropout \cite{gal2016dropout}, and evidential regression \cite{petit2004nonparametric}---within a crystal graph convolutional neural network architecture \cite{xie2018crystal} to predict adsorption energies in the OC20 dataset. The outputs are scored using a set of five diagnostic metrics: accuracy, sharpness, dispersion, calibration, and tightness, as shown in Figure~\ref{fig:uq_pipeline}.

\noindent This figure captures how UQ methods are commonly implemented in materials science studies ~\cite{shields2021bayesian}, and how stability (via convergence sensitivity tests with respect to loss hyperparameter $\lambda$ and ensemble size) and interpretability (with CGCNN as an accurate and interpretable architecture) are incorporated into a clear approach for building trust in the study outcomes. 

\begin{figure}[H]
\centering
\captionsetup{font=footnotesize}
\includegraphics[width=0.80\textwidth]{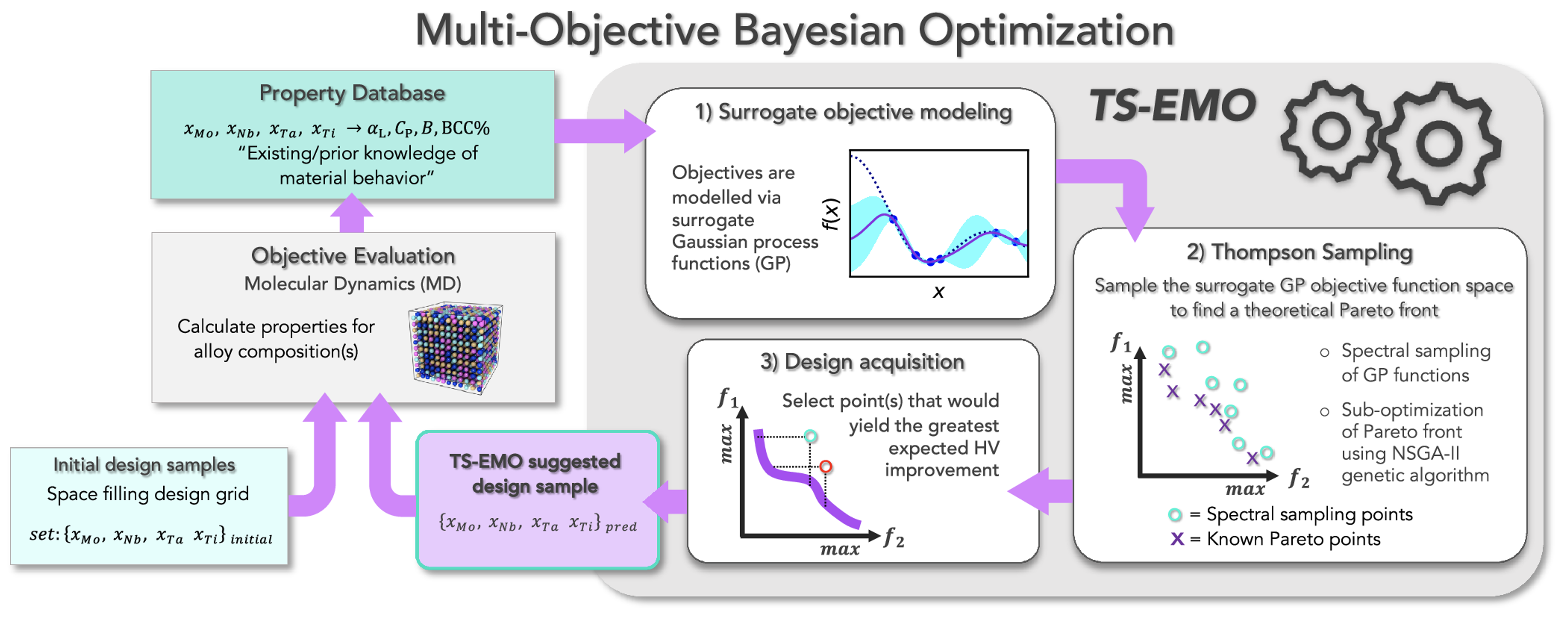}
\caption{Workflow for multi-objective Bayesian optimization of high-entropy alloys using TS-EMO, illustrating interpretable surrogate modeling, explainable decision acquisition, and generalizability across compositional space. Reproduced with permission~\cite{startt2024bayesian}. Copyright 2024 Nature Publishing Group UK London.}
\label{fig:TS-EMO}
\end{figure}

While handling uncertainty is key, interpretability and explainability also shape how predictions guide real-world materials design. Startt et al. \cite{startt2024bayesian} explore this through a multi-objective Bayesian optimization framework for finding optimal high-entropy alloy compositions (Figure \ref{fig:TS-EMO}). The goal is simple: search composition space efficiently, make the trade-offs visible, and check the best picks with an independent physics model.

\noindent TS-EMO supports TAI interpretability through Gaussian process surrogates that make model behavior legible. Explainability comes from the acquisition logic and Pareto visualization, which show why a composition is selected and what is gained along the front. The workflow also shows stability: repeated optimizations converge to similar Pareto fronts and hypervolume values. Robustness is suggested by consistent trends across MD and DFT, even without explicit noise tests.

\begin{figure}[H]
\centering
\captionsetup{font=footnotesize}
\includegraphics[width=0.78\textwidth]{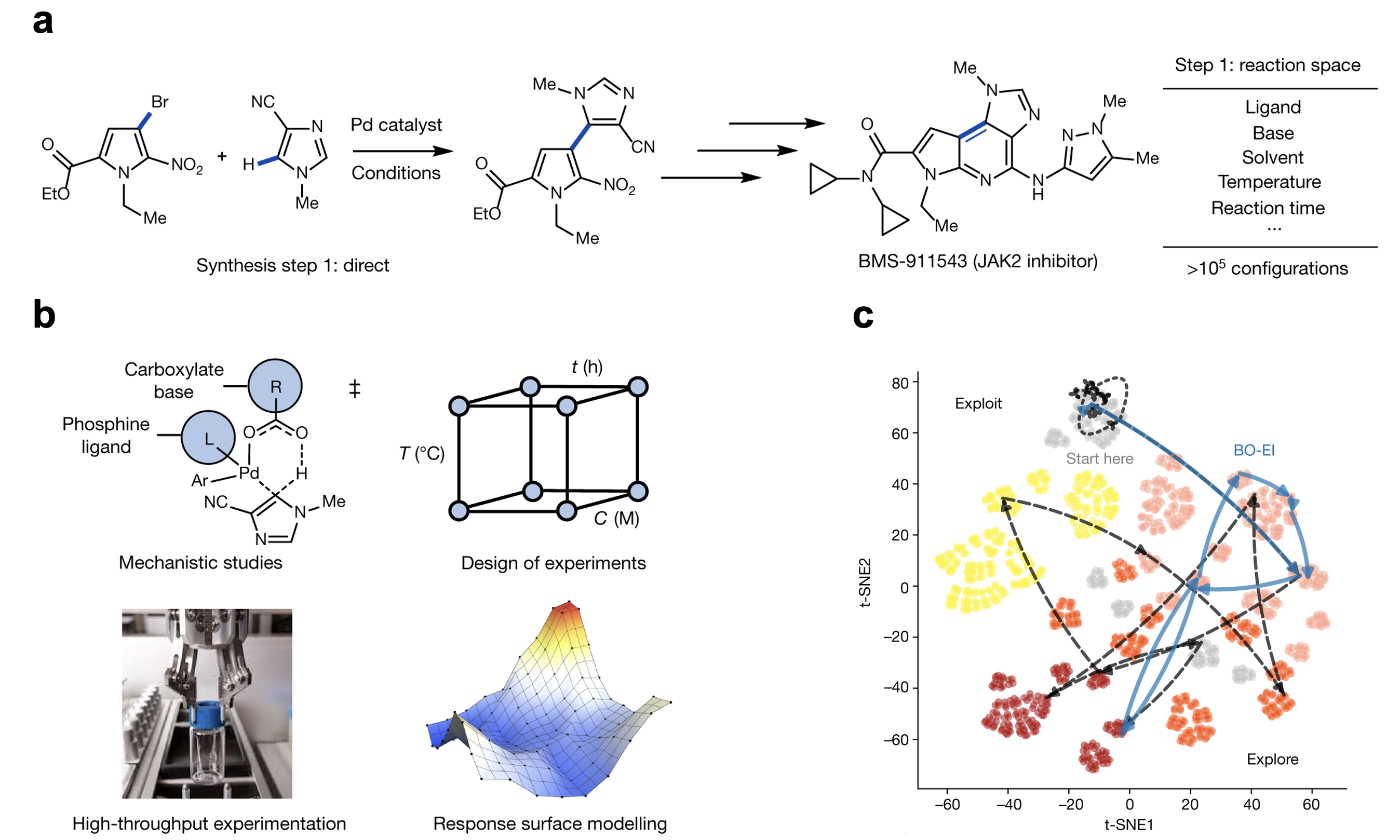}
\caption{Summary of the EDBO framework: (a) reaction space and synthetic complexity for a multistep inhibitor target; (b) mechanistic input, experimental design, and surface modeling; and (c) visualization of optimization trajectory within a high-dimensional design space, balancing exploration and exploitation. Reproduced with permission~\cite{shields2021bayesian}. Copyright 2021 Nature Publishing Group UK London.}
\label{fig:EDBO}
\end{figure}

In chemical synthesis, experimental design via Bayesian optimization (EDBO) framework explored large reaction spaces with over 100,000 configurations \cite{shields2021bayesian}. Synthesis planning improved when combining model-based optimization, automated experiments, and mechanistic insight (Figure \ref{fig:EDBO}).

\noindent Mechanistic variables and reaction-specific features improve interpretability. Using t-distributed stochastic neighbor embedding (t-SNE) projections to visualize model decisions across the design space and differentiate between exploration and exploitation behavior improves explainability.  The paper also speaks to robustness and stability. Repeated runs and alternative encodings show low variance and low worst-case loss, and similar outcomes from different starts.

\begin{figure}[H]
\centering
\captionsetup{font=footnotesize}
\includegraphics[width=0.85\textwidth]{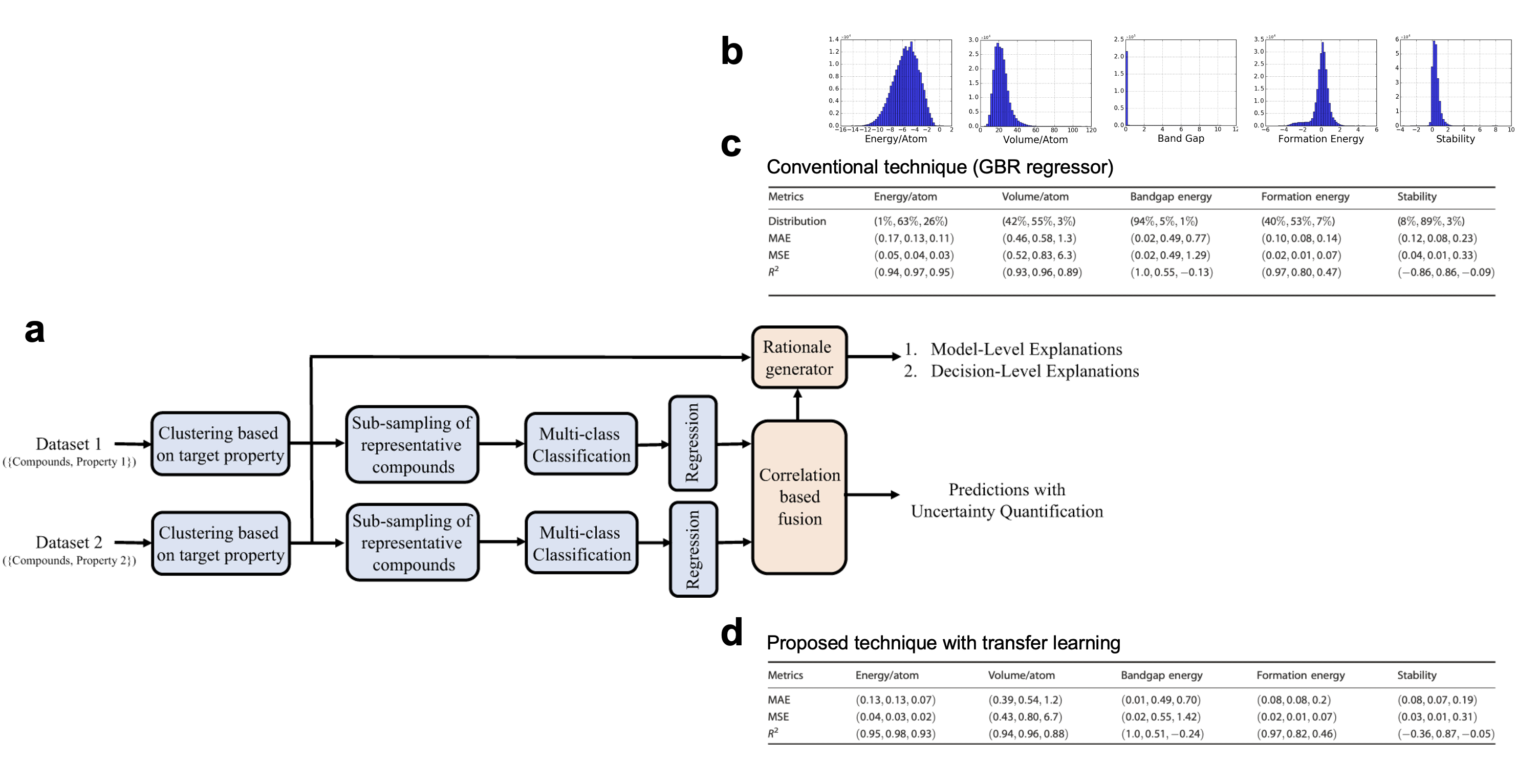}
\caption{Workflow and results from the reliable and explainable machine-learning pipeline for accelerated materials discovery: (a) end-to-end process showing class partitioning, balanced training, per-class modeling, and rationale generation; (b) histograms of property distributions, highlighting data imbalance; (c) class-specific performance metrics for a baseline gradient-boosted regressor, revealing weaker predictions for minority classes; and (d) performance gains from the proposed method with transfer learning, showing the largest improvements in minority classes. Reproduced with permission~\cite{kailkhura2019reliable}. Copyright 2019 Nature Publishing Group UK London.}
\label{fig:ReliableMLNew}
\end{figure}

While the previous example focused on chemical synthesis, Kailkhura et al.~\cite{kailkhura2019reliable} build a reliable and explainable pipeline for materials data that emphasizes fairness. It tackles skewed datasets, makes decisions traceable, and checks confidence. Figure \ref{fig:ReliableMLNew} puts the pieces together.

\noindent This study touches on many parts of trustworthy AI. In addition to fairness, interpretability comes from using separate models for each class instead of one large, opaque model. The rationale generator and clear routing steps make the decision process easier to follow, improving explainability. Robustness is strengthened by adding uncertainty estimates and a trust score that flags predictions in risky regions. Finally, stability shows up in the consistent results seen under different thresholds and cross-validation folds. 

\begin{figure}[H]
\centering
\captionsetup{font=footnotesize}
\includegraphics[width=0.80\textwidth]{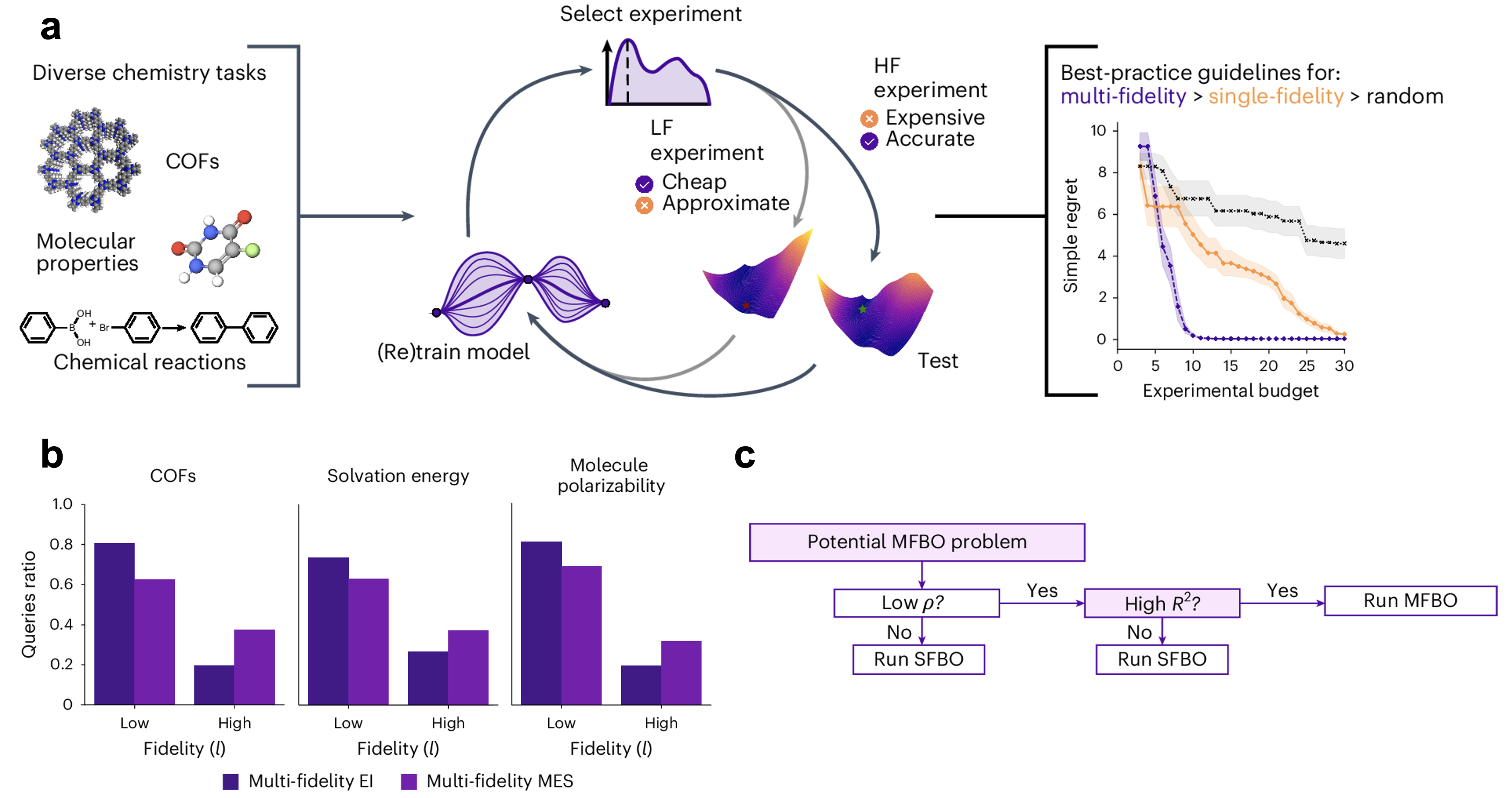}
\caption{Overview of multi-fidelity Bayesian optimization (MFBO) strategy: (a) the iterative selection process between low-fidelity (LF) and high-fidelity (HF) experiments, followed by model retraining; (b) simple regret curves showing MFBO outperforming single-fidelity and random strategies across budget; (c) ratio of LF to HF queries used across three chemistry tasks; and (d) a decision guide recommending MFBO when the LF source is inexpensive and sufficiently predictive. Reproduced with permission~\cite{sabanza2025best}. Copyright 2025 Nature Publishing Group US New York.}
\label{fig:mfbo}
\end{figure}

Bayesian methods can also support faster discovery by improving how experiments are planned across different levels of accuracy and cost. Sabanza et al.~\cite{sabanza2025best} explored this in a multi-fidelity optimization framework for materials and molecular discovery. The method balances high-fidelity (costly and accurate) with low-fidelity (inexpensive and approximate) experiments under resource constraints. Figure~\ref{fig:mfbo} illustrates the rule-based decision chart, principled kernel design, and an iterative selection mechanism to decide when multi-fidelity approaches are advantageous.

\noindent This study addresses several core TAI attributes. Stability comes from careful hyperparameter choices, including kernel scaling factors \( l_0 = \frac{2}{3} \) and \( l_1 = \frac{1}{3} \), which preserve meaningful cross-fidelity correlations. Robustness is evident in consistent performance across different LF cost and quality settings. Interpretability relies on simple regret metrics, structured decision rules, and clear kernel formulations. Explainability benefits from visual summaries that link performance to cost and fidelity use.

\begin{figure}[H]
  \centering
  \captionsetup{font=footnotesize}
  \includegraphics[width=0.85\textwidth]{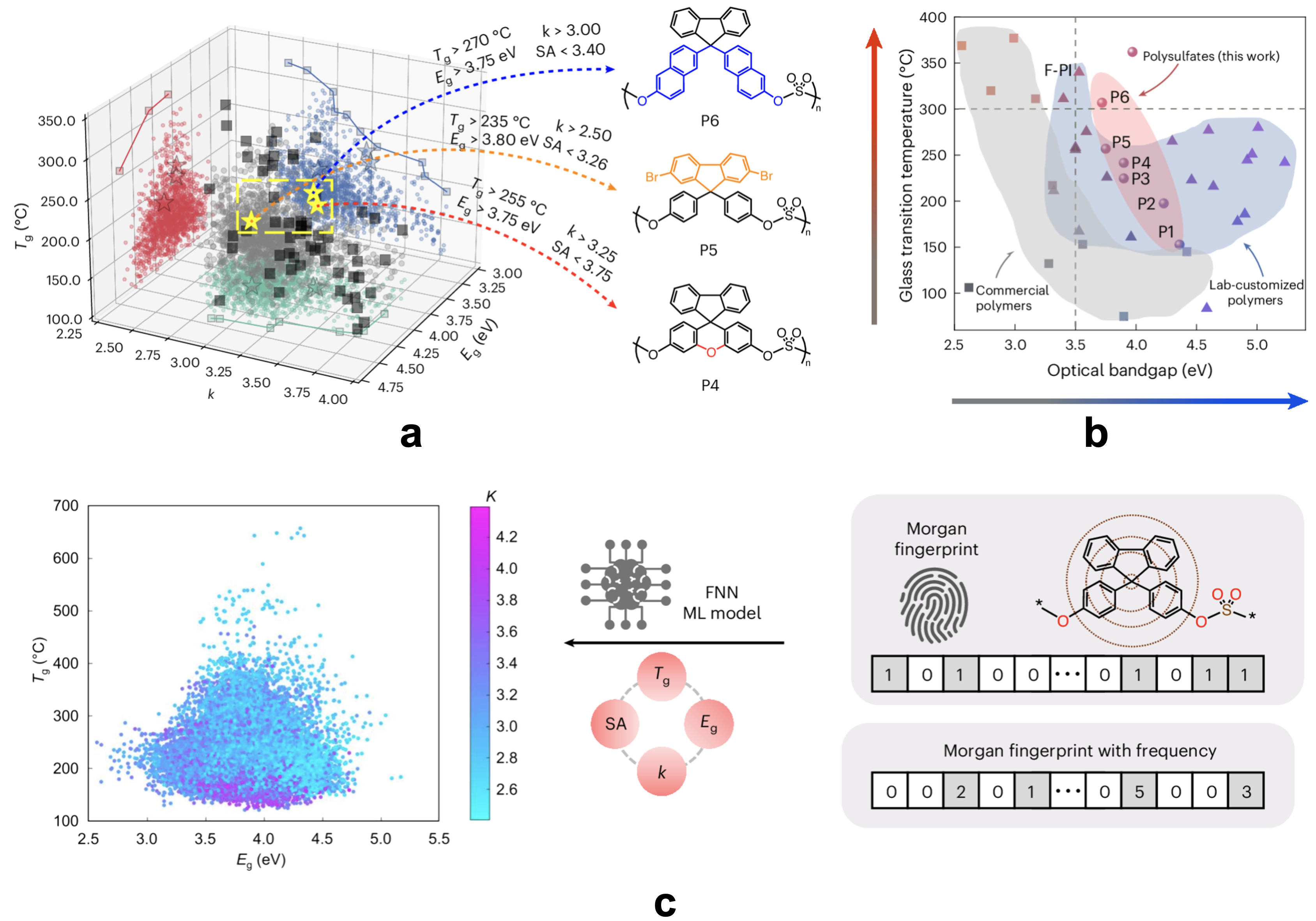}
  \caption{Machine learning-guided screening and experimental validation of heat-resistant polysulfates for electrostatic energy storage: (a) three-dimensional design space of predicted $T_g$, $E_g$, and dielectric constant $k$ used to select candidates P4, P5, and P6; (b) comparison of these candidates against commercial and lab-customized dielectric polymers; and (c) illustration of the FNN model architecture and input features, including Morgan fingerprint and Morgan fingerprint with frequency to explore the transition temperature, $T_g$ versus electronic bandgaps, $E_g$, for various dielectric constants, $k$, across the full polymer library. Reproduced with permission~\cite{li2025machine}. Copyright 2025 Nature Publishing Group.}
  \label{fig:nonBayesian_FNN}
\end{figure}

An example of TAI principles implemented in high-throughput screening study with non-Bayesian modeling was reported by~\citet{li2025machine} and visualized in Figure \ref{fig:nonBayesian_FNN}. The dataset was de-biased by resampling, introducing fairness. Two independently-trained ensembles demonstrated stability. The feed-forward neural network (FNN) models in this work are not interpretable according to the definition of \S \ref{sssec:interpretability}, SHAP~\cite{lundberg2017shap} analysis was introduced for explainability. 

A unique approach to non-Bayesian TAI was used when identifying a polarization switching mechanism for ferroelectric thin films~~\cite{griffin2020better}. Explainable methods such as k-means clustering were combined with fair data approaches to remove outlier and noisy data that would otherwise bias the model. The dimensionality reduction technique implemented an interpretable model with two hyperparameters, and a model stability study found that the predicted switching mechanism depended on the hyperparameter selection.

\begin{figure}[H]
  \centering
  \captionsetup{font=footnotesize}
  \includegraphics[width=0.90\textwidth]{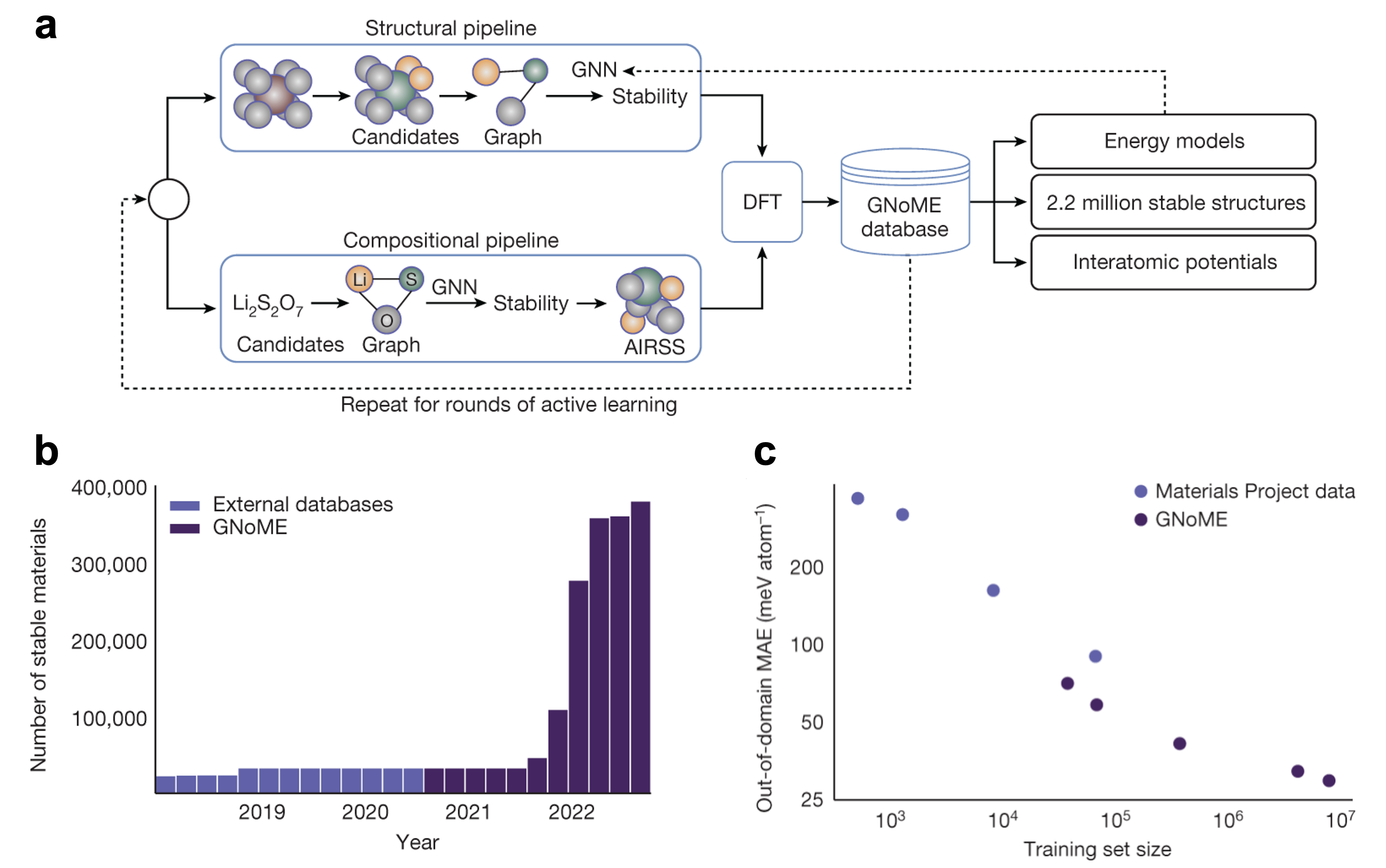}
  \caption{A summary of the GNoME-driven discovery process illustrates how model-based filtering and DFT validation form: (a) data flywheel that continuously enhances prediction accuracy; (b) the exploration powered by GNoME has resulted in the identification of 381,000 new stable materials—nearly an order of magnitude more than previously reported efforts; and (c) the framework demonstrates emergent generalization capabilities when evaluated on out-of-domain structures generated via random structure search, highlighting GNoME’s potential as a universal energy prediction model. Reproduced with permission~\cite{merchant2023scaling}. Copyright 2023 Nature Publishing Group.}
  \label{fig:scalability}
\end{figure}

Another example of comprehensive TAI is demonstrated by~\citet{merchant2023scaling}, where graph-based models were tasked with predicting new compounds (Figure \ref{fig:scalability}).
\noindent Deep ensembles indicated stability, and uncertainty quantification used the median rather than the mean statistic. Fairness was addressed through an unique encoding scheme that prevented identical samples from accidentally appearing in both training and testing data splits. Robustness was evaluated by training the model on data sampled at $T=400K$, then evaluating the model on data sampled at both $T=400K$ and $T=1000K$. 

\subsection*{Summary of GIFTERS Analysis}

These studies show that both Bayesian and non-Bayesian approaches contribute to TAI in materials science by integrating data-driven models with physics-based insights, offering scalable alternatives to probabilistic methods. However, some common trends were found in the works reviewed using the GIFTERS framework. 

First, most works prioritized a few aspects of TAI and omitted others; this is captured by an average GIFTERS score of 4.96/7 and a median score of 5 as shown in Table~\ref{table:1}. In some cases, omitting an aspect of TAI (i.e., interpretability for large neural network models) was justifiable, and indicates a need for new and better methods. However, an average of 4.96 indicates that most works omitted at least two of the seven principles included in the GIFTERS framework. In addition to four materials science studies that included all GIFTERS principles~\cite{wang2023bayesian, li2023critical, oviedo2019fast, deng2025machine}, the following section presents a study from climate research that demonstrates a comprehensive approach to TAI in other scientific domains as an example of how other domains are addressing this challenge.

A second trend observed concerns \textit{transparency}. In this review, if authors fully or partially released open source software to support their published results, their work received a ``1'' for \textit{transparency} to emphasize that the authors viewed transparency as important. However, software quality and associated metrics were not reported in any study, and standard software engineering methods such as unit testing for correctness were rarely implemented. This omission speaks directly to the reproducibility crisis in scientific results~\cite{semmelrock2025reproducibility, raff2025machine}. Expecting materials data-scientists to test and document their software is reasonable, in the same way that implementing calibration standards and reporting calibration results for experimental equipment is standard practice. Metrics for quantifying and reporting AI/ML software quality are leveraged in other domains, and we discuss these methods in the following section.

Like transparency, \textit{generalizability} of machine learning models and methods was emphasized in the works reviewed. However, there was less clarity when reporting the difference between generalization for \textit{interpolation} tasks versus \textit{extrapolation} tasks. This distinction is important to scientific AI/ML; physical models that capture fundamental characteristics and mechanisms are expected to generalize from one dataset to the next~\cite{wang2024extrapolation}. Metrics to evaluate differences in data distributions exist (e.g., earth mover distance~\cite{boudaa2020using}, Jensen-Shannon divergence~\cite{englesson2021generalized}), and in the following section we present studies where these metrics are leveraged to add trustworthiness to statements regarding generalizability.  

Differences appear between how trustworthiness is reported for Bayesian and non-Bayesian methods in materials science; we first discuss work using Bayesian methods. Bayesian work rightfully focuses on correctly quantifying uncertainty and estimating the posterior. The main risk is learning the wrong posterior. For example, data augmentation can change the effective likelihood, priors and likelihoods can be misspecified, and approximate inference can introduce bias~\cite{mena2021survey}. These risks are supported by data from full batch Hamiltonian Monte Carlo (HMC) studies~\cite{izmailov2021bayesian}. This connects directly to fairness. An imbalanced materials science dataset will still yield a confident posterior that reflects the imbalance \cite{kumagai2022effects}. Reporting uncertainty, physics-informed modeling, or implementing priors does not remove data bias. Instead, active strategies like stratified sampling, rebalancing, or fairness constraints during training are necessary to address fairness.

In non-Bayesian ML for materials science, model stability studies may be de-prioritized in favor of reporting high accuracy or novel discoveries. Testing stability often means repeating the training process for a model with different hyperparameters~\cite{ferrari2021troubling}, increasing required compute resources. Without this additional evaluation, what looks solid may in fact be fragile and the true strength of the model remains unknown~\cite{fengler2021non, myung2022challenges}. The complexity of machine learning optimization means small changes lead to different results, so that hyperparameter optimization is de-emphasized. Only when it is evident that results are not dependent on a single selection of hyperparameters can one have confidence in these models.

\section{Trustworthy AI Methods in the Broader Community}
\label{sec:what_other_people_do}

Trustworthy AI is no longer optional, especially in areas where decisions have real-world impact. Earlier in this review, we explored how methods contribute to building trustworthy AI/ML systems for materials discovery. We also identified key gaps in how some attributes are handled; broadly speaking, there is a strong (but inconsistent) reliance on statistical methods for evaluating models, a deficiency when considering methods that quantify transparency, and a tendency to consider reporting a few principles of trustworthiness sufficient. Because the materials science community has already indicated it considers TAI important, there is a need for methodologies and best-practices that support trustworthiness.

Importantly, these challenges are not unique to materials science. Other domains such as healthcare, environmental science, and public safety are confronting similar issues and have reported practical approaches to address them. Their experiences offer valuable insights that can be adapted and extended to the needs of materials discovery. In the following sections, we identify works with methods that may transfer well to materials science problems and illustrate how trustworthiness is discussed in a broader context.

\subsection*{Comprehensive Trustworthiness}

\begin{figure}[H]
\centering
\captionsetup{font=footnotesize}
\includegraphics[width=0.75\textwidth]{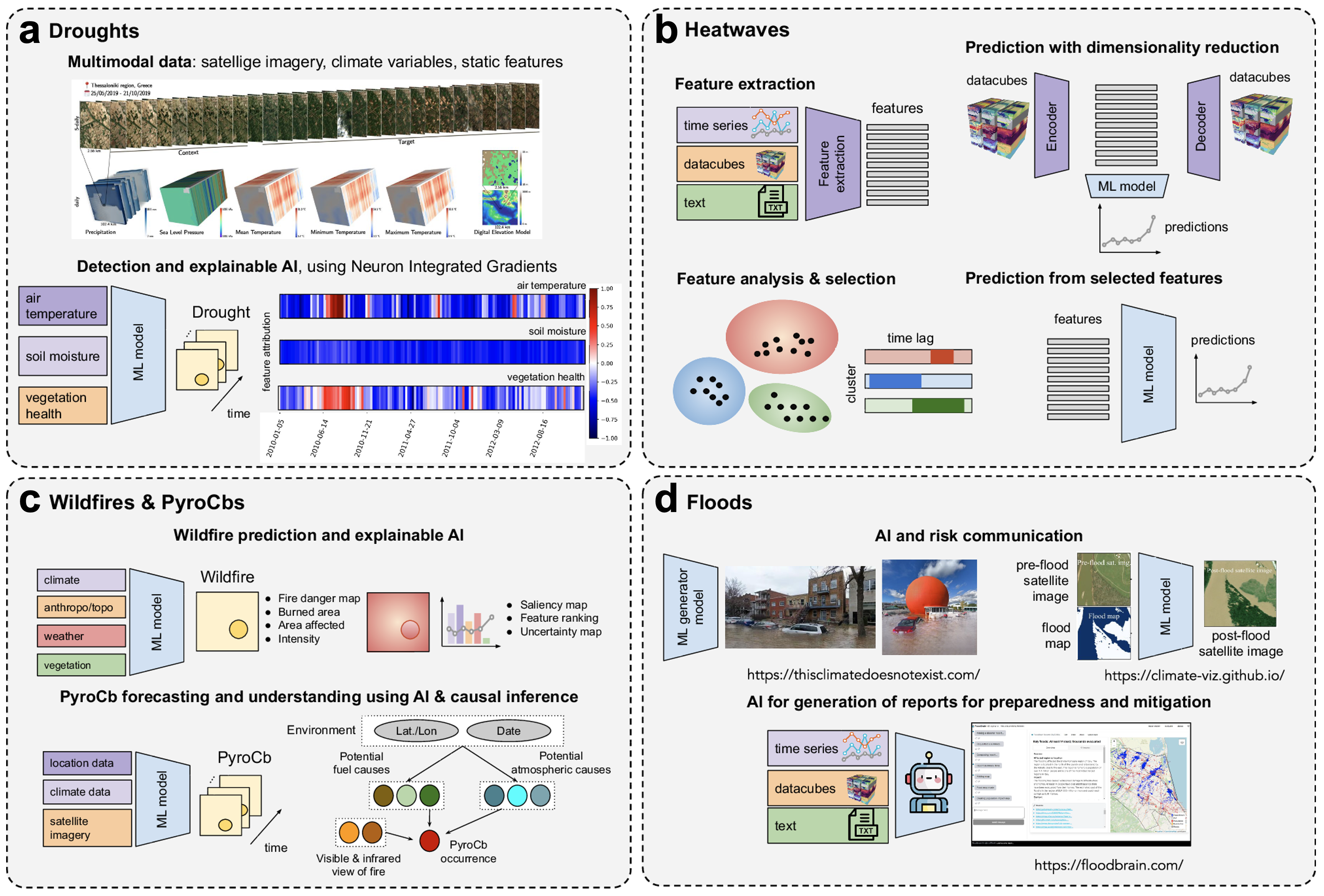}
\caption{Overview of AI methods applied to extreme climate events with emphasis on trust elements: (a) drought analysis integrates explainable AI with multimodal data and feature attribution; (b) heatwave prediction uses interpretable feature engineering and dimensionality reduction; (c) wildfire and PyroCb forecasting combines explainable models with causal inference; and (d) flood modeling illustrates risk communication and AI-driven report generation for preparedness and response. Reproduced with permission~\cite{camps2025artificial}. Copyright 2025 Nature Publishing Group UK London.}
\label{fig:TAI_climate}
\end{figure}

A strong example of comprehensive trustworthiness from outside materials science comes from climate research. A recent review showed how AI is used to detect, predict, and explain extreme events (e.g., floods, heatwaves, wildfires, and droughts) (Figure \ref{fig:TAI_climate}). The work addresses core TAI principles through methods familiar to materials science such as SHAP values and gradient maps, but also leveraged potentially-valuable techniques. For example adapting prototype layers in neural networks \cite{li2018deep} that use real-world examples which closely match a prediction could help material scientists trace specific outcomes back to microstructural features or chemical compositions. Another technique, counterfactual reasoning~\cite{verma2024counterfactual}, explored what might have happened under different conditions such as what outcome results if a heatwave had not been driven by greenhouse gases. Counterfactual approaches for materials science are being developed~\cite{teufel2023megan} but not widely implemented; this could enable researchers to test ideas before running expensive experiments or simulations, e.g., whether material behavior might alter under different conditions. 

\noindent The review also highlighted the role of communication in building trust, relating aspects of transparency, explainability, and generalizability to broader TAI ideas of human alignment and accountability. Interfaces such as warning dashboards, simplified summaries generated by language models, and standard report formats help public safety officials understand predictions and uncertainties quickly. In materials science, making results easier to understand could support human-in-the-loop analysis, leading to better human accountability and checks for human alignment. 

\subsection*{Fair Data Practices}

Building on the climate science example, we now turn to another domain where trust in AI/ML is not optional: healthcare. A recent paper by \cite{lai2025multimodal} introduces a multimodal artificial intelligence for ventricular arrhythmia risk stratification (MAARS), a multimodal model designed to predict sudden arrhythmic death in patients with hypertrophic cardiomyopathy. What makes MAARS stand out is not just its performance, but how it actively addresses key elements of TAI including fair data practices. 

\begin{figure}[H]
\centering
\captionsetup{font=footnotesize}
\includegraphics[width=0.75\textwidth]{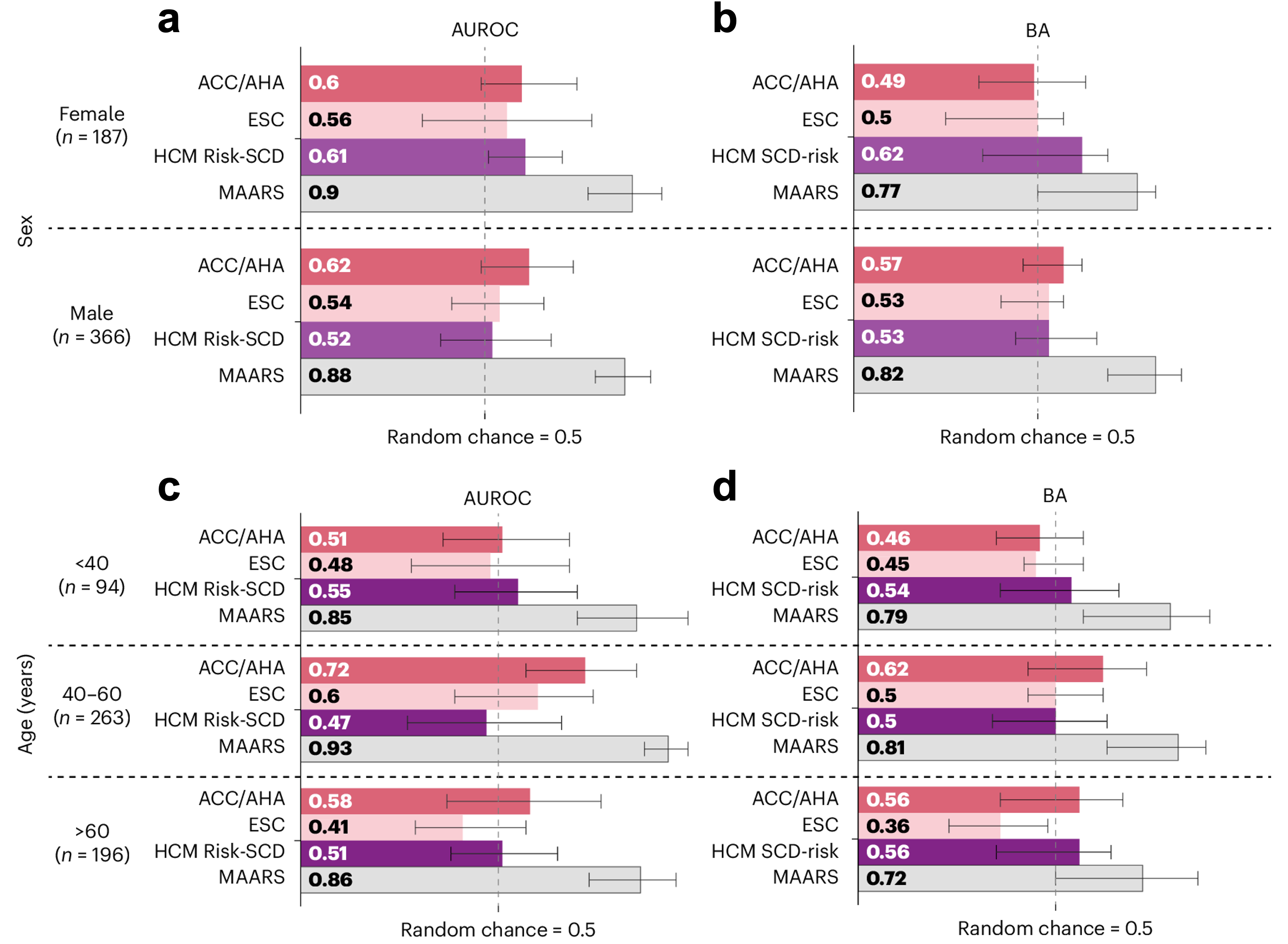}
\caption{Subgroup performance analysis of the MAARS model versus clinical risk guidelines in predicting sudden cardiac death among hypertrophic cardiomyopathy patients. Panels (a) and (b) show AUROC and balanced accuracy across sex, while (c) and (d) present the same metrics across three age categories. Reproduced with permission~\cite{lai2025multimodal}. Copyright 2025 Nature Publishing Group UK London.}
\label{fig:TAI_bio}
\end{figure}

\noindent In this work, the authors reported balanced accuracy within 0.05 and stable area under the receiver operating characteristic curve (AUROC) across age and sex groups. In addition to methods already discussed such as SHAP and attention maps \cite{roy2020attention}, a key method used in this work was the mid-fusion transformer, which let different data types (e.g., images, reports, and clinical records) interact within the model by creating a separate model branch for each kind of data; this addressed dataset fairness by reducing dataset bias while preventing overfitting due to reduced variance. Materials research often combines microscopy, diffraction, and composition data, so multimodal data fusion could be useful.

The criminal justice field provides another example of the need for fair data practices. A recent study introduced the recidivism clustering network (RCN), a framework that reduces bias and improves trust in risk predictions \cite{cavus2025transparent}. The approach combines multiple methods: class balancing (SMOTE), clustering methods (k-means, principal component analysis, t-SNE), deep learning, and SHAP explanations to assess recidivism. High accuracy was emphasized, along with decisions that can be explained, reviewed, and aligned with legal standards. Fairness in this study addressed data and features. For example, class imbalance was corrected with SMOTE so that minority cases were not lost during training. Class proportions were maintained constant in training and validation using a stratified 80/20 split. The model’s feature usage was evaluated using SHAP to ensure predictions were not influenced by proxy variables. Complementing this, clustering with k-means, PCA, and t-SNE helped check whether particular groups were over- or under-represented. These methods can be turned into practical steps for fair data use in materials science. For instance, rare phases or underrepresented processing routes can be balanced by using oversampling or augmentation. Instead of sampling artifacts, stratified splits guarantee that reported performance accurately represents the distribution of materials. Once bias is found, the fairness issue can be resolved by adjusting the feature, reweighting the data, and repeating the process. Fairness is incorporated into the model's interpretation of the data to minimize imbalance and proxy effects at their origin.

\begin{figure}[H]
\centering
\captionsetup{font=footnotesize}
\includegraphics[width=0.75\textwidth]{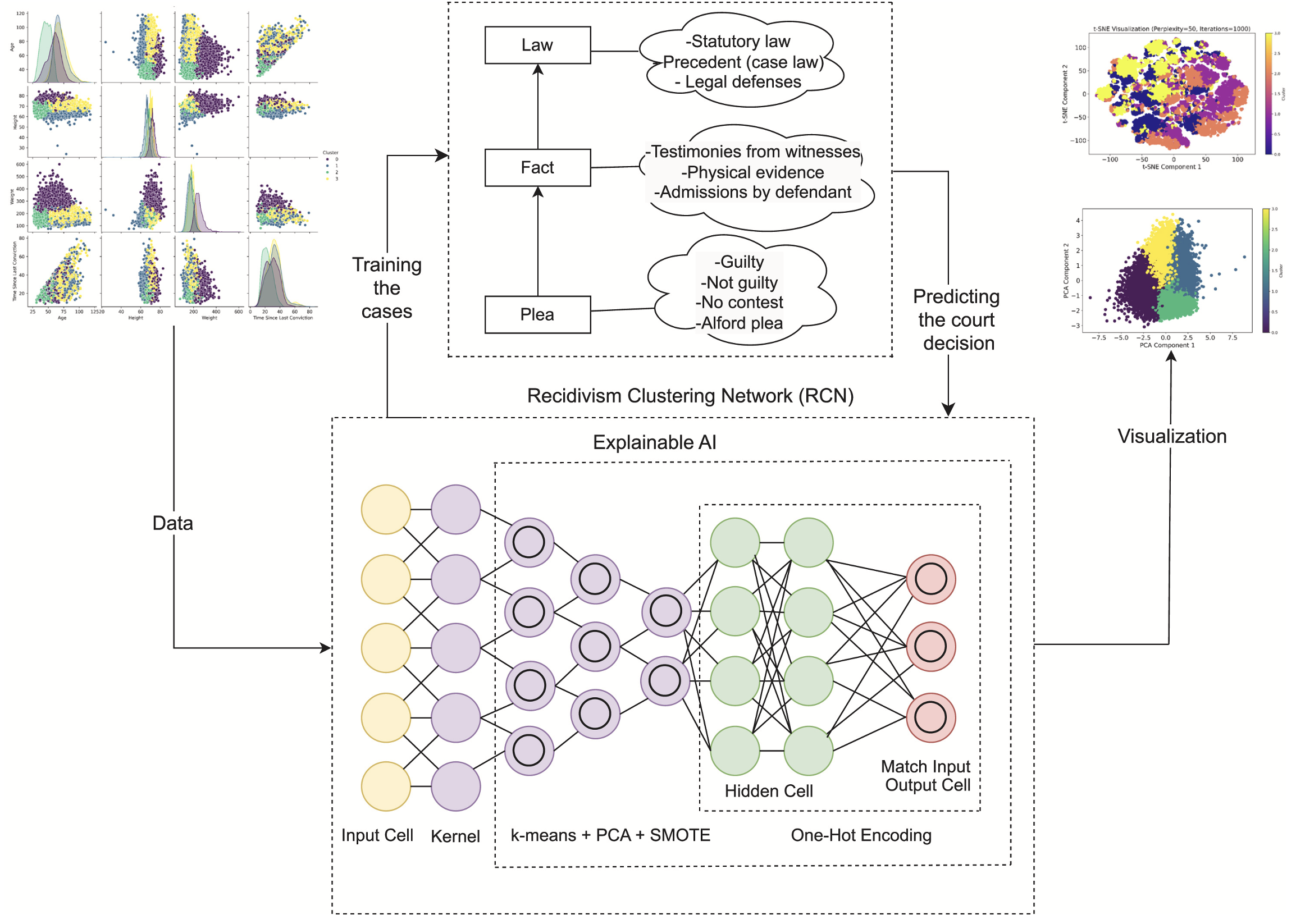}
\caption{Overview of the RCN framework for criminal justice applications. The pipeline integrates case data, data rebalancing (synthetic minority over-sampling technique), clustering, deep neural modeling, and SHAP-based explainability to support fair and interpretable court decision predictions. Visualizations at left and right show input feature distributions and learned offender clusters. Reproduced with permission~\cite{cavus2025transparent}. Copyright 2025 Elsevier.}
\label{fig:TAI_criminal}
\end{figure}

\begin{figure}[H]
\centering
\captionsetup{font=footnotesize}
\includegraphics[width=0.75\textwidth]{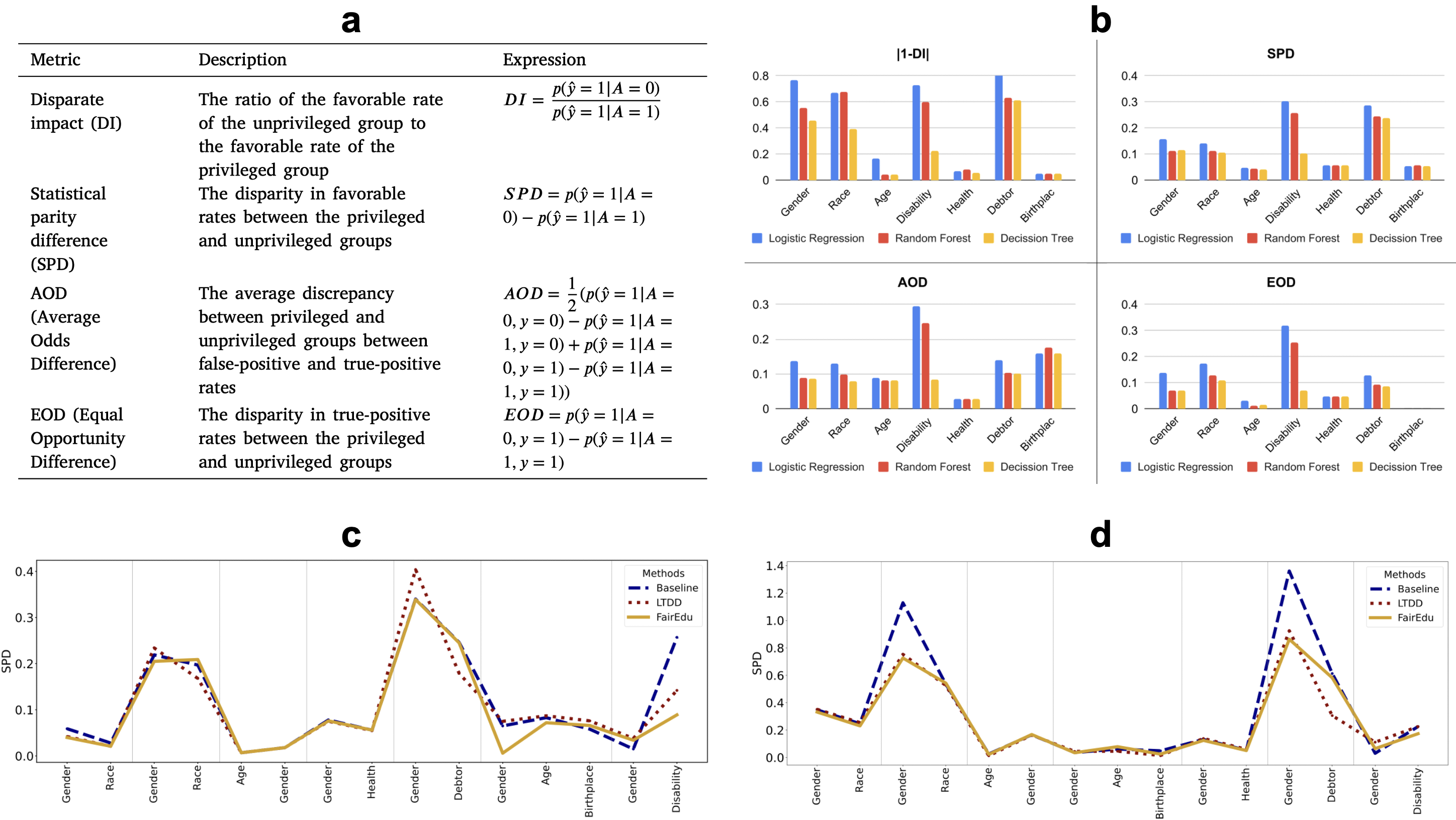}
\caption{Fairness evaluation of machine learning models in educational applications: 
    (a) key fairness metrics; (b) fairness results showing how each model handles bias for attributes like gender, race, age, disability, health, debtor status, and birthplace; (c) SPD values across sensitive features when using the baseline, LTDD, and the new FAIREDU method; and (d) SPD results across datasets, showing that FAIREDU improves fairness compared with baseline and LTDD while preserving accuracy. Reproduced with permission~\cite{pham2025fairedu}. Copyright 2025 Elsevier.}
\label{fig:TAI_educational}
\end{figure}

Educational AI is another high-stakes setting for data fairness because model decisions can shape access, achievement, and long-term opportunities for students. Pham et al.~\cite{pham2025fairedu} strengthen trustworthiness by moving fairness upstream into the data itself. Their method offers a practical way to create more equitable educational analytics by reducing group disparities without compromising accuracy on both public and institutional datasets. (Figure \ref{fig:TAI_educational}).

Fairness in this study is addressed at the data stage, before any model is trained. The authors present FAIREDU, a pre-processing technique that eliminates statistical connections between sensitive (e.g., age, gender, or race) and non-sensitive (e.g., math score) variables in order to improve feature representation. This is achieved through multivariate regression on the training set applied to the test set, preventing downstream models from learning proxy signals. The repaired data is evaluated with standard fairness metrics (Figure \ref{fig:TAI_educational}a). Because the method is model-agnostic, it can be compared fairly against a wide set of baselines, including fair-SMOTE, reweighing (RW), disparate impact remover (DIR), and linear-regression-based training data debugging (LTDD). 

These steps map directly to fair-data practices in materials science. In this case, sensitive attributes could be the instrument type, synthesis route, or lab or site of origin. To eliminate hidden proxies, regression-based correction may be applied to compositional features or microstructural descriptors. To make sure that evaluations represent actual distributions rather than sampling bias, stratified k-fold splits across phases, composition ranges, or processing routes are used. Reporting fairness metrics across significant subgroups and comparing them to resampling or reweighting techniques can convey trustworthiness. Collectively, the main takeaway is fix the features first, then use metrics to confirm.

\subsection*{Quantifying Generalizability}

Generalization can be achieved by using various methods during training the model, including data manipulation, representation learning, and learning paradigms~\cite{hwang2024rc, volkova2023overview, wang2022generalizing, jiang2022generalized, tang2022attainability, tran2022pruning}. Generalization is ideally evaluated by quantifying model performance on out-of-distribution (OOD) test datasets~\cite{li2025probing, li2024out}. Different measures such as the Wasserstein distance and Mahalanobis distance can be used to measure the difference between the data distribution used during training and the OOD data distribution used during testing~\cite{phan2022deepface, kwon2023improving}. Additional methods for quantifying OOD-ness include embedding trajectory for generative language models~\cite{wang2024embedding}, and considering which features are used to generate a prediction~\cite{sun2022dice}.

\begin{figure}[H]
\centering
\captionsetup{font=footnotesize}
\includegraphics[width=0.75\textwidth]{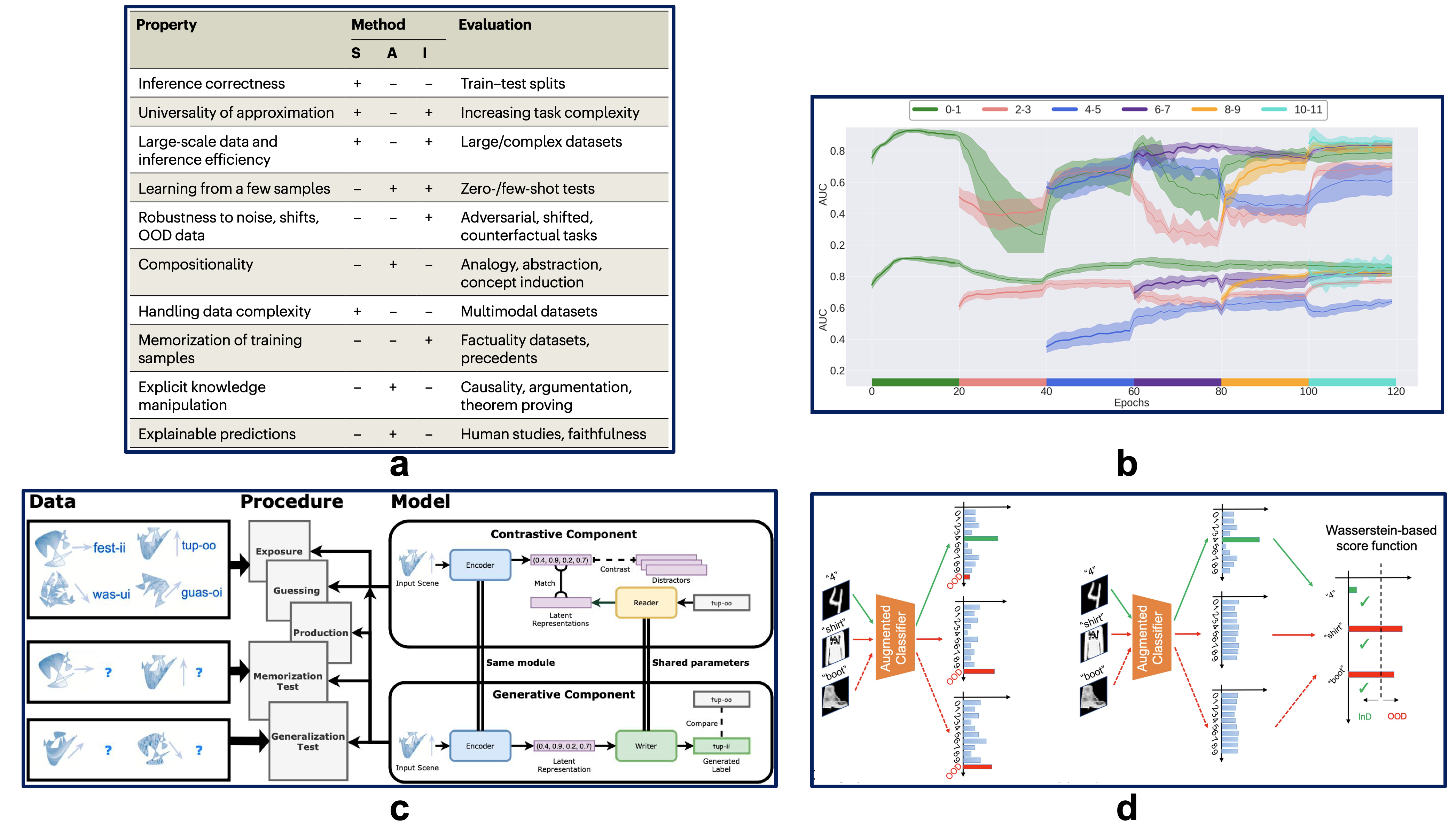}
\caption{Quantify generalizability across AI systems: (a) summarizes an evaluation taxonomy that links desired properties (e.g., few-shot learning, robustness to shifts/OOD data, compositionality, explainability) to appropriate testing protocols~\cite{Ilievski2025Aligning}; (b) continual-learning performance traces (AUC vs. training epochs) across sequential tasks, illustrating forgetting and recovery under shifts in class, time, modality, and institution~\cite{kiyasseh2021clinical}; (c) a compositional-generalization setup: data flow and procedure (exposure, guessing, production, memorization, generalization) alongside a model with shared encoders and contrastive/generative components used to measure production similarity, generalization score, and convergence on unseen combinations~\cite{galke2024deep}; and (d) Wasserstein-based OOD detection (WOOD), where an augmented classifier uses a Wasserstein score to separate in-distribution from OOD samples, enabling thresholded decisions and standard OOD metrics~\cite{wang2023wood}. Reproduced with permission~\cite{Ilievski2025Aligning}. Copyright 2025 Nature Publishing Group. Reproduced with permission~\cite{kiyasseh2021clinical}. Copyright 2021 Nature Publishing Group. Reproduced with permission~\cite{galke2024deep}. Copyright 2024 Nature Publishing Group. Reproduced with permission~\cite{wang2023wood}. Copyright 2023 IEEE.}
\label{fig:Gen_diffpaper}
\end{figure}

Researchers have begun to tackle generalization errors due to domain shift, continual learning, and scaling by developing ways to test and track generalization (Figure \ref{fig:Gen_diffpaper}).

One study shows that Wasserstein-based detection can be used to guide training so that in-distribution samples collapse toward one-hot outputs while out-of-distribution samples spread toward uniform predictions~\cite{wang2023wood}. This setup makes it easier to produce reliable AUROC scores for classification tasks. Another study in continual learning with cardiac signals introduces metrics like backward transfer, its time-dependent variants, and forward transfer~\cite{kiyasseh2021clinical}. These measures capture both how much knowledge is forgotten and how quickly models adapt and demonstrate that task curricula based on similarity improve both accuracy and retention. Using convergence measures, generalization scores, and production similarity, another study on compositional generalization demonstrates when models recombine known elements to handle unseen cases~\cite{galke2024deep}. A broader perspective suggests that real progress will come only when evaluation methods move beyond numbers and begin to capture human expectations~\cite{Ilievski2025Aligning}. In order to identify out-of-distribution data, facilitate incremental learning, and strike a balance between systematic reasoning and scalable training, \citet{Ilievski2025Aligning} suggest compositional stress testing, instance-based retrieval, and neurosymbolic techniques.

These developments provide useful strategies for materials science to quantify generalizability. Out-of-distribution scoring can be modified for use with diffraction, spectroscopy, or microscopy data. When new chemistries, phases, or processing routes appear, continual-learning metrics provide a check if models retain what they have learned. Structure-aware checks can be used on compositional and microstructural descriptors to see if models are recombining known elements in a systematic way when predicting the properties of new materials. These approaches can move the field away from ad hoc performance reports toward the more rigorous measure of generalizability essential for TAI in accelerated materials discovery.

\subsection*{Advancing Transparency in AI/ML Methods}

Transparency in model structure, parameters, data, and software has been proposed as a partial solution to the reproducibility crisis~\cite{semmelrock2025reproducibility, raff2025machine} in scientific machine learning, and is therefore a central part of TAI. In the works reviewed, materials science researchers addressed transparency by releasing model weights and an accompanying code repository. However, reproducibility as discussed by reproducibility researchers~\cite{raff2025machine} also encompasses repeatability, replicability, adaptability, model selection, data quality, incentive for scientific rigor, and maintainability. Full transparency requires effort beyond open source code. 

This broader understanding of transparency is evident in the literature, where expectations for transparent AI/ML vary among disciplines. In a pure AI/ML context, ~\citet{semmelrock2025reproducibility} identify seven artifacts which must be shared for an experiment to be ``complete'': hypothesis, prediction method, source code, hardware specifications, software dependencies, experiment setup, and experiment source code. In healthcare, checklists have been proposed to evaluate whether the use of AI/ML in a published work is suitable and transparent~\cite{scott2021clinician, cabitza2021need, el2023reporting}, with the International Journal of Medical Informatics now requiring submissions to self-evaluate against the checklist~\cite{cabitza2021need} before submission.  

On a more granular level, software engineering metrics may be used to quantify the  readability of the source code~\cite{posnett2011simpler, rahman2025tool}; readability aids in transparency by making the code easier to understand. In the earth sciences and social sciences, transparency in data processing is addressed by standardizing how data labeling is documented~\cite{wirz2024increasing}; this is distinct from fair data processes because it examines how the methods are reported and not just which methods are implemented.  

A transparency-oriented framework \textit{FAIR}~\cite{wilkinson2016fair, wilkinson2018design} has been proposed to standardize data science stewardship. Importantly, the FAIR framework has been connected to evaluation metrics \cite{wilkinson2018design}, introducing a quantitative approach to transparency that is useful for the scientific community. Finally, the Foundation Model Transparency Index~\cite{bommasani2023foundation} evaluates each model across thirteen aspects of transparency. This granularity ensures repeatable results and post-hoc analysis, and empowers materials science researchers that their work is both correct and reusable.

\begin{figure}[!h]
  \centering
  \captionsetup{font=footnotesize}
  \includegraphics[width=0.90\textwidth]{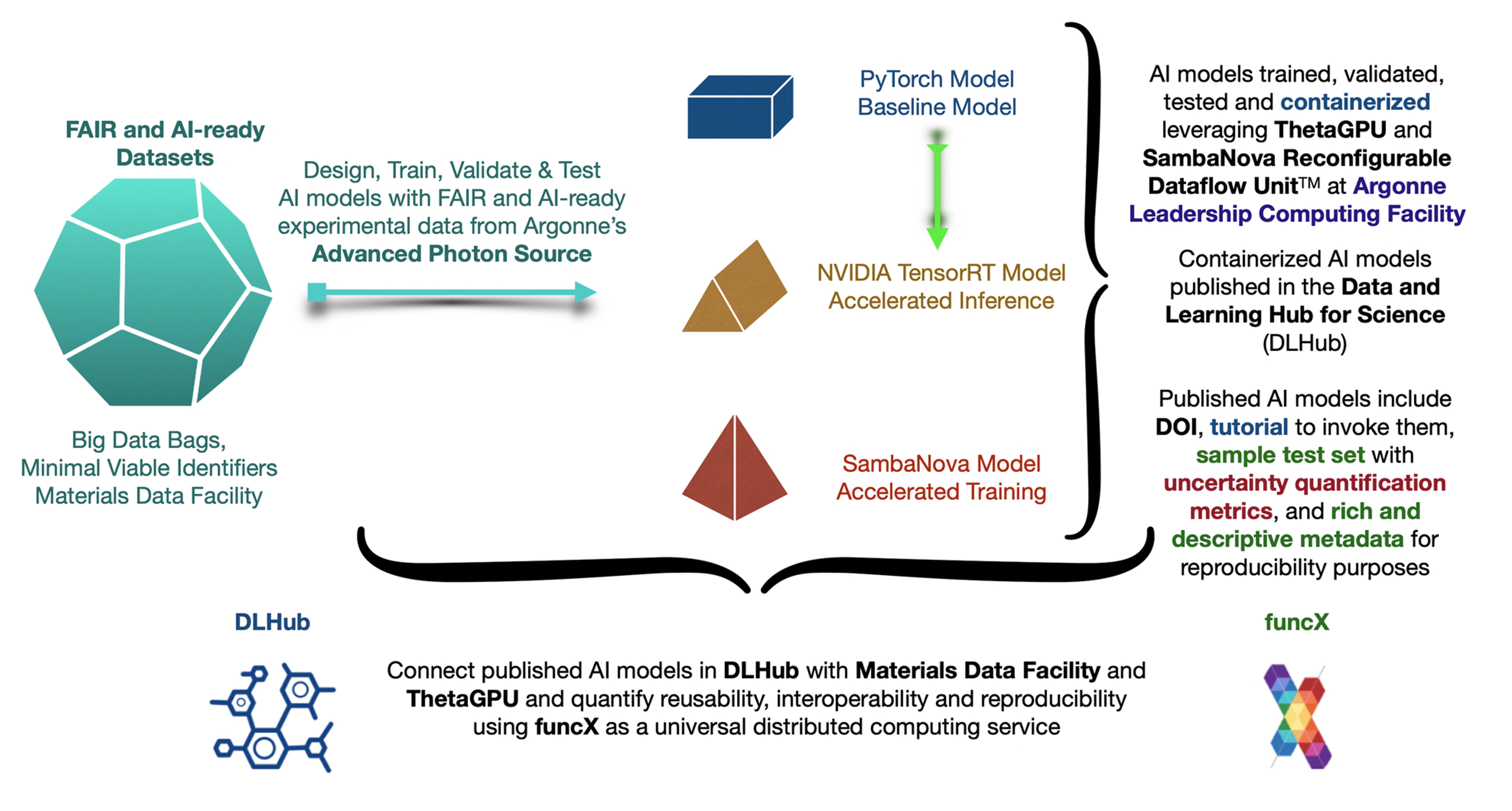}
  \caption{The FAIR-AI framework for accelerating high-energy diffraction microscopy through AI model development and deployment. The workflow integrates FAIR and AI-ready datasets from the Materials Data Facility with PyTorch, TensorRT, and SambaNova models, which are trained and containerized using ThetaGPU and ALCF resources. Models are published in DLHub with DOIs, tutorials, sample test sets, uncertainty quantification metrics, and rich metadata to ensure findability, accessibility, interoperability, and reusability. Reproduced with permission~\cite{ravi2022fair}. Copyright 2022 Nature Publishing Group UK London.}
  \label{fig:fairness}
\end{figure}

\subsection*{Reporting Stability Studies}

\begin{figure}[H]
\centering
\captionsetup{font=footnotesize}
\includegraphics[width=0.75\textwidth]{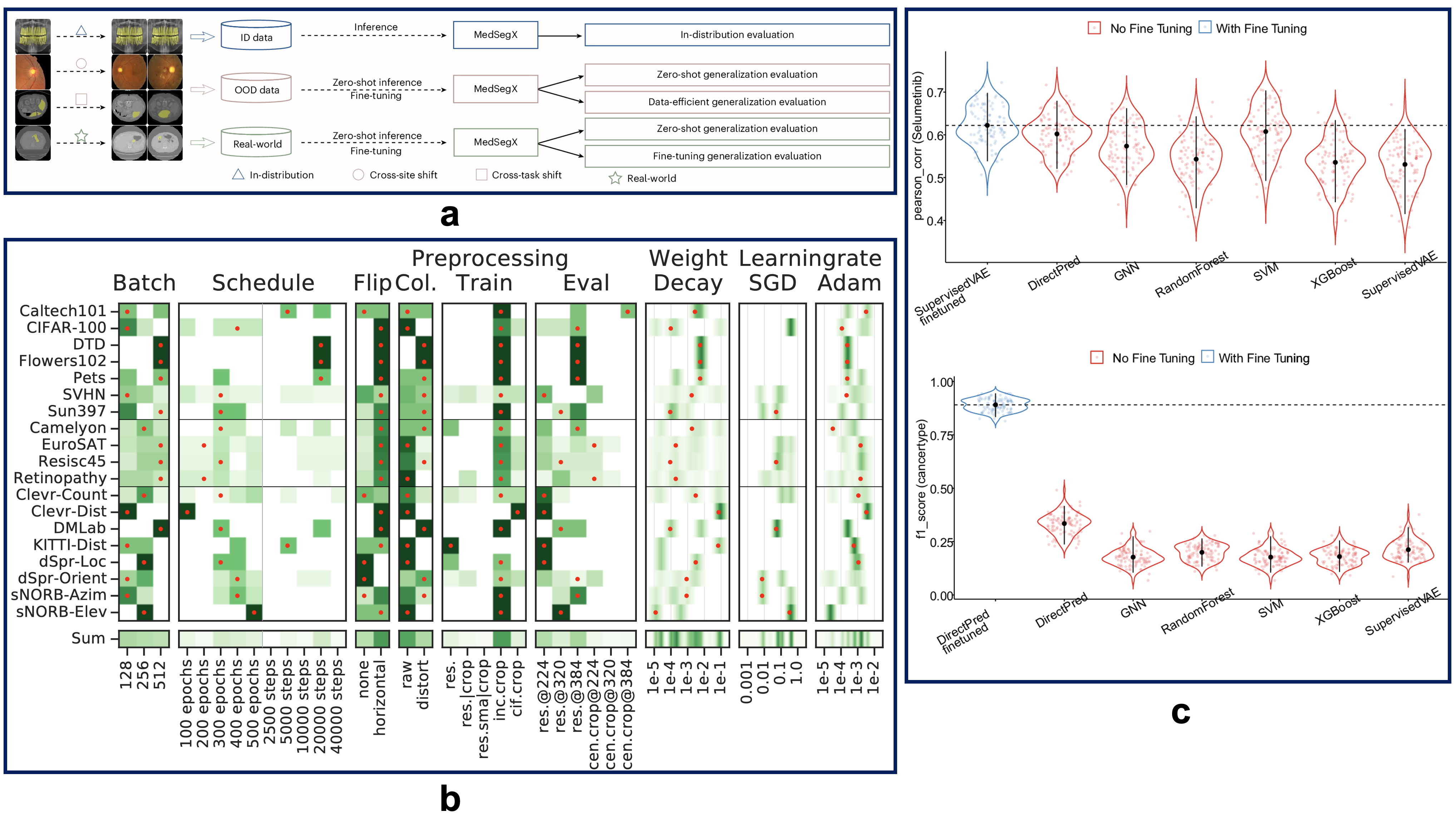}
\caption{Advancing stability reporting in AI/ML methods: (a) evaluation flow for the MedSegX generalist medical-image segmenter, outlining in-distribution tests, two OOD regimes (cross-site and cross-task), and a real-world track, with both zero-shot and data-efficient (few-label fine-tuning) assessments~\cite{zhang2025generalist}; (b) VTAB’s hyperparameter-sweep summary heatmap, which visualizes, per downstream dataset, the best setting (red dot) and the full set of near-optimal settings (green background)~\cite{zhai2019large}; and (c) Flexynesis violin plots showing how modest fine-tuning shifts performance distributions under matched versus shifted domains, with means and 95\% bootstrap CIs overlaid for deep-learning and classical baselines~\cite{uyar2025flexynesis}. Reproduced with permission~\cite{zhang2025generalist}. Copyright 2025 Nature Publishing Group. Reproduced with permission~\cite{zhai2019large}. Copyright 2019 arXiv preprint arXiv:1910.04867. Reproduced with permission~\cite{uyar2025flexynesis}. Copyright 2025 Nature Publishing Group.}
\label{fig:Transp_HPO}
\end{figure}

Reporting model stability goes beyond transparency because it includes an analysis of how hyperparameters affect model performance. As shown in Figure \ref{fig:Transp_HPO}, the best method of hyperparameter selection to promote stable modeling is an active area of research~\cite{karl2023multi, khan2025comparative, mehdiyev2024quantifying, moosbauer2021explaining}. For example, \citet{bischl2023hyperparameter} present six different algorithms for selecting hyperparameters and report that the impact of the hyperparameter on the model performance is ``subtle''.  Bayesian strategies for hyperparameters optimization include Gaussian process and ensemble methods~\cite{bergstra2011algorithms}, evolutionary methods such as covariance matrix adaptation (CMA)~\cite{hansen2001completely}, and bandit-style pruning such as Hyperband~\cite{li2018hyperband} or asynchronous successive halving (ASHA)~\cite{jamieson2016non}. An evolutionary algorithm tuned hyperparameters encoded as a chromosome~\cite{toderean2025heuristic}. An alternative to hyperparameter analysis involves perturbing the model parameters directly to see how this affects model performance~\cite{wen2020uncertainty, kim2021polymer}. Frameworks like Optuna~\cite{akiba2019optuna}, Vizier~\cite{golovin2017google}, SMAC~\cite{hutter2011sequential}, Hyperopt~\cite{bergstra2015hyperopt}, Katib~\cite{zhou2019katib}, and Tune~\cite{liaw2018tune} combine methods for ease of use.  

Reports highlight different aspects of model stability. The effect of compute budget on identifying optimal hyperparameters is addressed by the visual task adaptation benchmark (VTAB)~\cite{zhai2019large}. Another study standardizes training and evaluation, then eliminates important components to explain performance gains~\cite{zhang2025generalist}. One study compares hyperparameter tuning methods, then maps out Pareto fronts, runs extensive statistical tests, and reports efficiency~\cite{yu2025calibrating}. Flexynesis performs hyperparameter tuning, then summarizes outcomes across model families with paired statistical tests and an auditable dashboard~\cite{uyar2025flexynesis}. Finally, a Taylor diagram~\cite{opoku2024prediction, taylor2001summarizing} condenses hyperparameter effects to three statistical measures: the Pearson correlation coefficient $R$, the standard deviation of the set of target values $\sigma_z$, the standard deviation of the predicted values $\sigma_{\hat{z}}^2$ and the standard deviation of the error (SDE) so that $\text{SDE}^2 = \sigma_z^2 + \sigma_{\hat{z}}^2 - 2\sigma_z \sigma_{\hat{z}}R$. These practices clarify whether results come from the data, the architecture, or the tuning, and ensure that the method used to obtain the reported model is stable with respect to the search space, budget, selection rule, validation protocol, number of seeds, and statistical tests.

\section{Future Directions and Recommendations}
\label{sec:future_directions}

Progress in TAI depends on filling the gaps that benchmarking, case studies, and cross-domain research have only hinted at. Examples from multiple disciplines indicate that trust in AI is being built by researchers striving for more transparent, reliable, and explainable methods. This is important for materials science because we can now adapt proven methods to our own challenges and build trust into our models from the start. We can also go further, drawing inspiration from other fields to develop new AI/ML methods that simultaneously accelerate discovery and improve trustworthiness. We next focus on four fronts that will inspire the next wave of TAI experimental materials science and engineering research.

\begin{itemize}
    \item \textbf{Physics-informed and knowledge-guided learning when no closed-form model exists:} Much of today’s physics-informed work assumes some governing equations are known; this is largely untrue in experimental settings. Directly encoding weak physics~\cite{fajardo2023fundamental}, domain regularities~\cite{hasani2022closed},  and physics-inspired constraints~\cite{loh2022surrogate, kon2025exp, pestourie2023physics} into learning systems without the need for a closed-form model is a feasible next step. Lab protocols and instrument transfer functions may be treated as AI/ML model priors, with ablation studies indicating how each prior changes calibration and error under shifting conditions. The next step is to test these ideas at the bench by planning measurements to deliberately stress physics-inspired constraints. This could lead to new trustworthiness metrics, such as the frequency with which a system violates a physical principle, or the degree of violation, and demonstrate when trust is improved through soft physics.
    
    \item \textbf{Trust for robotics AI/ML in self-driving labs:} Sophisticated closed-loop platforms exist to automate experiments; these systems may benefit from proposed requirements for trust in robotics~\cite{haskard2025secure,cappuccio2024autonomous}. A lab autonomy readiness scale has been proposed to provide a structured path for robotic experimentation. Each stage in the scale was defined by clear criteria. These requirements include the use of verified interlocks~\cite{leong2025steering}, exploration guided by uncertainty~\cite{xu2024uncertainty}, automatic delay when calibration drifts~\cite{doloi2025democratizing}, and reproducibility across various instruments~\cite{canty2025science}. To complement this structure, task level scorecards were expected to be created~\cite{tom2024self}. These comprehensive approaches satisify a need for auditability, and connects uncertainty with explanation in a practical approach. As a result, repeatable and verifiable processes strengthen confidence in robotic experimentation.

    \item \textbf{Using NLP models with explicit trust metrics:} Trustworthiness metrics used in natural language processing can be leveraged for materials discovery. Confidence metrics indicate when model outputs are insufficient for guiding experimental action and when suggestions should instead be withheld pending stronger evidence~\cite{fu2025deep}. Errors should be analyzed to ensure the model meets reliability standards~\cite{chen2025benchmarking}. Human scientists use documented logical processes and contextual awareness when progressing from literature to hypotheses; model alignment with expert human methods should be examined~\cite{cai2025natural} so that scientists can extend their expertise without compromising baseline reliability~\cite{Gao2025IncreasingAlignment}. Moreover, text mined candidates should be paired with physics-based descriptors and other interpretable rules, so that outputs can be cross-checked against established scientific principles prior to experimental validation~\cite{bai2025preferable}. The result is a pathway for trusting, reproducing, and adopting NLP model predictions with confidence that accelerates materials discovery.

    \item \textbf{Human-in-the-loop and domain expertise as first-class safeguards:} Human-in-the-loop methods and domain expertise can be brought together in a trust framework for materials discovery that integrates confidence gating, error auditing, expert-guided control, domain-preserving adaptation, and physically meaningful interpretability into one workflow. Progress proceeds from deep domain knowledge, so uncertainty-aware metrics enable human supervision that reduces bias and improves uncertainty estimates while still maintaining reliability~\cite{scheurer2025role}. In Bayesian optimization, for example, Gaussian process surrogates can offer calibrated uncertainty for acquisition decisions~\cite{biswas2024dynamic}. Next, human experts who can identify unexpected outcomes and fix malfunctions in sophisticated software systems ensure that errors missed by current AI or robotic systems are still addressed quickly~\cite{liu2025balancing}. From a systems design perspective, domain-adaptation is performed by data-informed scientists who adapt workflows and design spaces while preserving the trust signals that foundation models already provide; this can be achieved by combining evidence from multiple sources and levels of fidelity, including negative results. Guidelines for interpretable AI also recommend that human-alignment be planned in advance~\cite{mosqueira2023human}, and this entails establishing precise roles, plans, and deadlines that align with the expert audience's cognitive capabilities. 
\end{itemize}

\section{Conclusion}
%The use of artificial intelligence in materials discovery is changing how science gets done. This review highlights how AI—through machine learning, generative models, and active learning—is speeding up the search for new materials and cutting costs, especially in areas like batteries, catalysts, and advanced manufacturing. But speed is not the whole story. 
To make reliable discoveries using AI/ML, the materials science community currently emphasizes elements of trustworthiness summarized by the principles of the \textit{GIFTERS} framework.  These approaches are already helping scientists predict material behavior with greater accuracy and confidence. However, most materials science work leveraging AI/ML methods falls short of comprehensive trustworthiness, and may benefit from AI/ML trustworthiness methods used in other disciplines such as healthcare and climate science. Finally, an analysis of future directions in materials science research demonstrates the need to design systems that actively involve human experts, follow physical laws, adapt to new data, and explain their reasoning. This is not just about making better predictions, it is about building AI that scientists can trust to help solve real-world problems.

\subsection*{Author Contributions}

A.S.D, B.A., J.H.-S. conceived and designed the project. A.S.D., B.A. reviewed the literature and drafted the manuscript. J.H.-S supervised the project. A.S.D., B.A., S.K., and J.H.-S. discussed the results. A.S.D., B.A., S.K., and J.H.-S. reviewed and edited the manuscript. All authors contributed to the manuscript preparation.

\subsection*{Funding}

A.S.D. acknowledges supports from the University of Toronto’s Eric and Wendy Schmidt AI in Science Post-doctoral Fellowship, a program of Schmidt Sciences. 

\subsection*{Competing Interests}

All authors declare no financial or non-financial competing interests. 

{
    \small
    \bibliography{ref}
}

\end{document}